\documentclass[]{aa}

\usepackage{longtable}
\usepackage{threeparttablex}
\usepackage{graphicx}
\usepackage{longtable}
\usepackage{multicol}
\usepackage{supertabular}

\usepackage{txfonts}
\newcommand{\kms}{\,km\,s$^{-1}$}

\newcommand{\ind}{IN\,Com}
\newcommand{\inc}{IN\,Com }
\newcommand{\Halpha}{H$\alpha$ }
\newcommand{\Halphad}{H$\alpha$}

\begin{document}

   \title{Surface magnetic activity of the fast-rotating G5-giant IN\,Comae, central star of the faint planetary nebula LoTr~5\thanks{Based on data obtained with the STELLA robotic observatory in Tenerife, an AIP facility jointly operated by AIP and the Instituto de Astrofisica de Canarias, and on data acquired with PEPSI using the Vatican Advanced Technology Telescope (VATT) jointly operated by AIP and the Vatican Observatory.}}

\author{Zs.~K\H{o}v\'ari\inst{1}
\and K.~G. Strassmeier\inst{2}
\and K.~Ol\'ah\inst{1}
\and L.~Kriskovics\inst{1}
\and K.~Vida\inst{1}
\and T.~A. Carroll\inst{2}
\and T.~Granzer\inst{2}
\and I.~Ilyin\inst{2}
\and J.~Jurcsik\inst{1}
\and E.~K\H{o}v\'ari\inst{3}
\and M.~Weber\inst{2}
}

\offprints{Zs. K\H{o}v\'ari}

\institute{Konkoly Observatory,
Research Centre for Astronomy and Earth Sciences, Hungarian Academy of Sciences,
Konkoly Thege \'ut 15-17., H-1121, Budapest, Hungary; 
  \email{kovari.zsolt@csfk.mta.hu}
  \and Leibniz-Institute for Astrophysics Potsdam (AIP), An der Sternwarte 16,
D-14482 Potsdam, Germany
\and E\"otv\"os University, Department of Astronomy, Pf. 32., H-1518, Budapest, Hungary
}

   \date{Received ; accepted}

\abstract  
{On the asymptotic giant branch, low to intermediate mass stars blow away their outer envelopes, forming planetary nebulae. Dynamic interaction between the planetary nebula and its central progenitor is poorly understood. The interaction is even more complex when the central object is a binary star with a magnetically active component, like it is the case for the target in this paper. }
{We aim to quantify the stellar surface activity of the cool binary component of \ind\ and aim to explain its origin. In general, we need a better understanding of how central binary stars in planetary nebulae evolve and how this evolution could develop such magnetically active stars like \ind . }
{We present a time-series of 13 consecutive Doppler images covering six months in 2017 and use it to measure the surface differential rotation with a cross-correlation method. Hitherto unpublished high-precision photometric data from between 1989 to 2017 are presented. We apply Fourier-transformation based frequency analysis to both photometry and spectra. Very high resolution ($R\approx$200,000) spectra are used to update \ind's astrophysical parameters by means of spectral synthesis. }
{Our time-series Doppler images show cool and warm spots coexisting with an average surface temperature contrast of $-1000$\,K and $+300$\,K with respect to the effective temperature. Approximately 8\%\ of the stellar surface is covered with cool spots and $\approx$3\%\ with warm spots. A consistent cool polar spot is seen in all images. The average lifetime of the cool spots is not much more than a few stellar rotations (one month), while the warm spots appear to live longer (3 months) and are mostly confined to high latitudes. We found anti-solar surface differential rotation with a shear coefficient of $\alpha=-0.026\pm0.005$ suggesting an equatorial rotation period of 5.973$\pm$0.008\,d.  We reconfirm the 5.9-day rotation period of the cool star from photometry, radial velocities, and \Halpha line-profile variations. A long-term $V$-brightness variation with a likely period of 7.2\,yr is also found. It  appears in phase with the orbital radial velocity of the binary system in the sense brightest at highest velocity and faintest at lowest velocity, that is, at the two phases of quadrature. We redetermine [Ba/Fe], [Y/Fe] and [Sr/Fe] ratios and confirm the overabundance of these s-process elements in the atmosphere of \ind.}
{}%


\authorrunning{K\H{o}v\'ari et al.}
\titlerunning{Time-series Doppler imaging of IN\,Com}

\maketitle


\section{Introduction}\label{intro}

Undergoing the asymptotic giant branch (AGB) evolution phase, low to intermediate mass stars blow away their outer envelopes, forming planetary nebulae that will be ionized by the developing white dwarfs in their center. However, interaction between the planetary nebula and its central progenitor is far from understood. The scheme is even more complicated, when the central object is a binary star, which may be the majority \citep[e.g.,][]{2004ApJ...602L..93D,2017NatAs...1E.117J}. When the binary system is embedded in a common envelope, the components evolve together, as initially proposed by \citet{1976IAUS...73...75P}. Inside the envelope the stellar cores spiral together and finally they may merge, forming a fast-rotating giant star, such as FK\,Comae stars \citep[for a recent overview of binary mergers see the contribution by Ph. Podsiadlowski in][]{2014apa..book.....G}. However, if the envelope material is ejected before merging, the binary evolution may end up in a system involving a white dwarf and a main sequence companion. In this case some of the ejected nebular material is captured by the companion, yielding cataclysmic variability, while the accretion gradually spins up the accreting star. In addition, a powerful magnetic dynamo can also develop, since the common envelope is largely convective and supposed to rotate differentially \citep{2003ASPC..293..100T}. However, the theory of common envelope evolution is extremely complex, involving different physical processes (e.g., ejection, accretion, spiralling, magnetic braking) on different time-scales (thermal, dynamical, hydrodynamical, magneto-hydrodynamical, etc.), therefore a self-consistent, all-comprehensive treatment is still not feasible \citep[cf.][]{2001ASPC..229..239P}.

In this paper we revisit one of our former Doppler imaging targets, IN\,Comae \citep[=HD\,112313,][hereafter Paper~1]{1997A&A...322..511S}, the central object of the faint planetary nebula LoTr~5 \citep{1980MNRAS.193..521L}. The system consists of a white dwarf and a giant G-star, which is evident from the composite IUE spectra  \citep{1983ApJ...269..592F}. Actually, the central object is a long-period ($P_{\rm orb}\approx2700$\,d$\approx7.4$\,yr) binary system, consisting of a  rapidly rotating magnetically active G5 giant ($v\sin i=67$\,kms$^{-1}$) and a hot ($T_{\rm eff}\approx150\,000$\,K) subdwarf companion \citep[cf.][and their references]{2017A&A...600L...9J}.  It was even suggested that the G5 component had a close companion, forming a hierarchical triple system \citep{1987A&A...180..145J,1991AJ....101.2131M}. However, this option has recently been disproved by \citet{2014A&A...563L..10V}, who suggested that either the white dwarf (or hot subdwarf) formed a close binary together with a yet undiscovered star of some 2--3\,$M_{\odot}$, or alternatively the orbital plane was not coplanar with the waist of the bipolar nebula. The cool star was found to be barium-rich \citep{1997A&A...320..913T} as a consequence of contamination by s-process elements from the AGB progenitor envelope \citep[cf.][]{2011MNRAS.418..284B}, supporting the presumption of spinning-up of the cool component by accreting the ejected envelope material.

Rotational and orbital period determinations of \inc in the past yielded misinterpretations and even contradictory results. As fundamental period 5.9-day was reported by \citet{1989SvAL...15..149N}, which was attributed to rotation of the G5 giant (see Paper~1). However, \citet{1991AJ....101.2131M} argued for 1.2-day as the most prominent photometric signal (i.e., the $1-f$ alias of 5.9 days), and also for 1.75 days, based on radial velocity measurements, as a possible orbital period of the assumed close binary. \citet{1994A&A...286..211J} could not confirm this binary orbit, but they claimed that 1.2-day was indeed correct. After a revision, however, \citet{1996A&A...307..200J} found the 5.9-day period as more realistic. Oddly, \citet{1993AcA....43..445K} found an even shorter period of 0.25 days, which, however, has not been confirmed ever since. Over and above, the radial velocity measurements in Paper~1 did not support the close binary hypothesis. Only recently, \citet{2017A&A...600L...9J} have updated the first reliable orbital motion detection by \citet{2014A&A...563L..10V} and proposed an orbital period of 2717$\pm$63 days ($\approx$7.4 years) for the G5 star.

\inc has been observed in X-rays by \emph{XMM-Newton} with all three European Photon Imaging Cameras (EPIC) on 6 June 2002, and by \emph{Chandra} on 4 December 2002 \citep{2010ApJ...721.1820M}.  Spectral fitting shows that the X-ray emission is characterized by two components at about 0.65 keV (both \emph{XMM-Newton} and \emph{Chandra}) and at 2.27 and 3.49 keV (\emph{XMM-Newton} and \emph{Chandra}, respectively), while the X-ray luminosity was measured as $\log L_{\rm X}\approx30$\,erg\,s$^{-1}$. \citet{2012IAUS..283..204G} fitted the \emph{XMM-Newton} EPIC planetary nebula spectrum with 8$\times$10$^{30}$\,erg\,s$^{-1}$ at 0.61 and 3.1 keV, and constructed a spectral energy distribution model from the \emph{XMM-Newton} data, the available \emph{IUE} spectra, and ground based optical and near-infrared photometry. The non-local thermodynamic equilibrium (non-LTE) model showed good agreement with a contribution from a supposed G5\,III companion star. It seems, that the X-ray emission dominantly originates from the corona of the magnetically active late-type component of the binary \citep[see][for details]{2010ApJ...721.1820M}.

Our study may contribute in moving towards a better understanding how central binary stars in planetary nebulae evolve and how this evolution could develop such fast-rotating magnetically active stars like \ind.
The paper is organized as follows. In Sect.~\ref{obs} we present the photometric and spectroscopic observations. In Sect.~\ref{di} we provide updated astrophysical parameters for \inc and present a time-series Doppler imaging study. With that we analyze the spot evolution and measure the surface differential rotation. In Sect.~\ref{phot} we study the photometric and spectroscopic variability of the \inc system, while in Sect.~\ref{halp} the \Halpha behaviour is examined. The results are summarized and discussed in Sect.~\ref{disc}.

\section{Observations}\label{obs}

\subsection{Photometry}\label{obs_phot}

Most of the photometric observations were obtained with the T6 and T7 (`Wolfgang' and `Amadeus', respectively) \mbox{0.75-m} automatic photoelectric telescopes (APTs) located at Fairborn Observatory in southern Arizona  \citep{1997PASP..109..697S}, operated by AIP \citep{2001AN....322..325G}. Altogether 1364 data points were observed in Johnson $V$ (T6) and 943 in Str\"omgren $y$ (T7) colours between February 1996--June 2017 (JD\,2,450,117--2,457,911). Besides, 643 Johnson $V$ observations were collected with the \mbox{1-m} RCC telescope of Konkoly Observatory, Budapest, located at Piszk\'estet{\H o} mountain station, Hungary, between January 1989--June 1993 (JD\,2,447,530--2,449,141). The old photometric data from the literature completed with the new, yet unpublished $V$ and $y$ observations are plotted together in Sect.~\ref{phot} in the top panel of Fig.~\ref{figphot}.

\subsection{Spectroscopy}

Spectroscopic observations were carried out with the \mbox{1.2-m} STELLA-II telescope of the STELLA robotic observatory \citep{2010AdAst2010E..19S} located at the Iza\~{n}a Observatory in Tenerife, Spain. It is equipped with the fibre-fed, fixed-format STELLA Echelle Spectrograph (SES) providing an average spectral resolution of $R=55\,000$. Altogether 230 high-resolution echelle spectra were recorded between January 26 and June 23, 2017. The spectra cover the 3900--8800\,\AA\ wavelength range without gaps. Further details on the performance of the system and the
data-reduction procedure can be found in \citet{2008SPIE.7019E..0LW,2012SPIE.8451E..0KW} and \citet{2011A&A...531A..89W}. The average signal-to-noise ratio (S/N) of the spectra is 140:1. Table~\ref{Tab1} in the Appendix summarizes the division of the spectra into 12 independent subsets (dubbed S01--S12) which are used for Doppler imaging. 

In addition, 20 ultra-high resolution ($R=200\,000$) spectra were collected during March 03--15, 2017 with the \mbox{1.8-m} Vatican Advanced Technology Telescope (VATT) fiber linked to the Potsdam Echelle Polarimetric and Spectroscopic Instrument (PEPSI) at the nearby Large Binocular Telescope (LBT). PEPSI's characteristics and performance were described by \citet{2015AN....336..324S,2018A&A...612A..44S}. With cross disperser (CD) III in the blue arm and CD~V in the red arm the set-up provided a wavelength coverage of 4800--5440\,\AA\ and 6280--7410\,\AA , respectively. The 90-min exposures gave typical S/N of 100:1 for the red and 50:1 for the blue wavelength regions. This data set is used primarily to refine some of the fundamental astrophysical parameters of \inc (see Sect.~\ref{astrop}). The log for these observations is given in Table~\ref{TabVAT} in the Appendix.

\section{Doppler imaging}\label{di}

\subsection{Adopted stellar parameters}\label{astrop}

The effective temperature, the surface gravity, the metallicity, and the microturbulence velocity are re-examined by applying the spectrum-synthesis code SME \citep{2017A&A...597A..16P} to the ultra-high resolution PEPSI spectra. Our synthesis is based on MARCS model atmospheres \citep{2008A&A...486..951G} and assuming local thermodynamic equlibrium (LTE). Atomic parameters are taken from the Vienna Atomic Line Database \citep[VALD,][]{1999A&AS..138..119K}. For the spectrum synthesis we used the 4800--5441\,\AA\ and 6278--7419\,\AA\ wavelength ranges. SME is applied for all single PEPSI spectra and the individual results are combined in order to estimate their error bars. This way we get $T_{\rm eff}$= $5400\pm100$\,K, $\log g$=$2.6\pm0.1$, [Fe/H]=$-0.10\pm 0.05$ and $\xi_{\rm mic}$=$2.0\pm0.4$\,km\,s$^{-1}$  for the effective temperature, surface gravity, metallicity and microturbulence, respectively. We note that our temperature and surface gravity values are in good agreement with the recent result by \citet{2018MNRAS.476.1140A}.
The radial-tangential macroturbulence dispersion of $\approx$7\,km\,s$^{-1}$ is estimated according to \citet{1981ApJ...251..155G} and \citet{1986ApJ...310..277G}.  SME was used to measure the Ba, Y and Sr abundances \citep[cf.][]{1997A&A...320..913T} as well. For the abundance fits we kept all of the other redetermined astrophysical parameters fixed. We note that for the Ba line fit we take the average of 40 high quality STELLA spectra since the spectral gap between the blue and red arms of the PEPSI data highly overlaps with the 5519--6694\,\AA\  region of the 31 neutral and singly ionized barium lines taken from VALD. The new astrophysical data of \inc are summarized in Table~\ref{astropars}.

\begin{table}[h]
\caption{The astrophysical properties of IN\,Com} \label{astropars}
 \centering
 \begin{tabular}{lll}
 \hline
  \hline\noalign{\smallskip}
  Parameter               &  Value \\
  \hline
  \noalign{\smallskip}
  Spectral type            & G5\,III  \\
  Gaia distance [pc]    &  $506\pm12$\\
  $V_{\rm br}$    [mag]          & $8.69\pm0.03$ \\
   $(B-V)_{\rm HIP}$     [mag]         & $ 0.835\pm0.004$ \\
  $M_{\rm bol}$     [mag]        & $0.01\pm0.08$ \\
  Luminosity [${L_{\odot}}$]         & $78\pm6$ \\
  $\log g$ [cgs]        &          $ 2.6\pm0.1$ \\
  $T_{\rm eff}$ [K]           &           $5400\pm100 $    \\
  $v\sin i$ [km\,s$^{-1}$]           &         $67.0\pm1.5$ \\
  Photometric period [d]   &           $5.934\pm0.001 $ \\
  Equatorial rotation period [d]   &           $5.973\pm0.008 $ \\
  Differential rotation coefficient &           $-0.026\pm0.005 $ \\
  Inclination  [\degr]            &       $45\pm15$ \\
  Radius      [$R_{\odot}$]           &      $11.1^{+5.0}_{-2.2}$   \\
  Mass          [$M_{\odot}$]           & $1.8\pm 0.4$   \\
  Microturbulence  [km\,s$^{-1}$] & $2.0\pm 0.4$ \\
  Macroturbulence  [km\,s$^{-1}$] & 7.0 (adopted) \\
  Metallicity [Fe/H] &  $-0.10\pm 0.05$ \\
  Barium/iron ratio [Ba/Fe] &  $0.85\pm 0.25$ \\
  Yttrium/iron ratio [Y/Fe] &  $0.27\pm 0.12$ \\
  Strontium/iron ratio [Sr/Fe] &  $\gtrsim1.0$ \\
\hline
 \end{tabular}
\end{table}

The projected equatorial velocity of 67$\pm$1.5\kms\ \citep[Paper~1, but see also][]{2014A&A...563L..10V} and a  45$\pm$15\degr\ inclination (cf. Paper~1) together with the equatorial rotation period of 5.973\,d (cf. Sect.~\ref{rotmod}) yields a stellar radius of 11$^{+5.0}_{-2.2}$R$_{\odot}$. Together with $T_{\rm eff}$=5400$\pm$100\,K this radius is consistent with the G5\,III classification in the literature. 

The new \emph{Gaia} DR-2 parallax of 1.977$\pm$0.046\,mas \citep{2018arXiv180409365G} yields a distance of 506$\pm$12\,pc for \ind. The brightest ever observed $V$ magnitude of $8\fm69\pm0\fm03$ (see later Fig.~\ref{lcrad}), while neglecting any interstellar and circumstellar extinction \citep[cf][]{1999AJ....118..488C}, results in  an absolute magnitude $M_{V}$=$0\fm17\pm0\fm08$. This, together with a bolometric correction for a G5 giant of $BC=-0\fm163$ taken from \citet{1996ApJ...469..355F} gives a bolometric magnitude of $M_{\rm bol}$=$0\fm01\pm0\fm08$, and thus a luminosity of $L=78\pm6\,L_{\odot}$ in fair agreement with the value just from the Stefan-Boltzmann law, but with a much smaller error bar. Also, taking above radius and our measured gravity the stellar mass is $\approx 1.8\pm0.4$\,M$_{\odot}$.

\subsection{Definition of data subsets}

The spectroscopic data used for the Doppler-imaging (DI) process are all from the first half of 2017 and are distributed fairly uniformly over the five-months STELLA run. In spite of the relatively short rotation period of 5.9\,d, we still got satisfactory phase coverage with between 8 and 13 spectra per image for 12 subsequent intervals of typically one stellar rotation each. An additional subset can be formed from the available PEPSI spectra, with a pretty dense phase coverage. Table~\ref{disets} summarizes the timely distribution of the 13 data subsets (dubbed S01--S12 and P01 for the PEPSI spectra) while Tables~\ref{Tab1} and \ref{TabVAT} in the Appendix record the observing logs for the STELLA and the PEPSI spectra, respectively.

\begin{table*}[]
 \centering
\caption{Temporal distribution of the subsequent datasets for each individual Doppler image}
\label{disets}
\begin{tabular}{c c c c c c }
\hline
\hline\noalign{\smallskip}
Data & Mid-HJD & Mid-date & Number & Data range & Data range \\
subset &2\,450\,000+  & 2017+& of spectra &in days& in $P_{\rm rot}$  \\
\hline
\noalign{\smallskip}
S01  & 7784.490 & Jan-30  & 8 & 6.910 &  1.164  \\
S02  &   7791.670   & Feb-07 & 11 & 5.091 & 0.858 \\
 S03  &   7813.465 & Feb-28  & 13 & 7.984 & 1.345 \\
 S04 & 7819.742  &  Mar-07 & 11 & 4.967 & 0.837  \\
P01 & 7823.892 & Mar-11 & 18 & 8.141 & 1.372\\
S05  &  7826.080  & Mar-13 & 13 & 5.230 & 0.881\\
 S06  &  7835.673  & Mar-23 & 9 & 4.732 &  0.797\\
S07  &   7841.438 & Mar-28 & 13 &  5.050 & 0.851  \\
 S08  &   7864.277  & Apr-20 & 9 & 8.166   & 1.376 \\
S09 &  7877.389  &  May-03  & 9 & 10.149 &  1.710 \\
S10   &  7892.854  & May-19  & 11 & 5.214  &0.879  \\
S11  &   7898.804   & May-25 & 10 & 5.175 &  0.872 \\
S12  &   7912.724   & Jun-08  & 9 & 5.135 &  0.865 \\
\hline
\end{tabular}
\end{table*}

\subsection{Image reconstruction with iMap}\label{imap}

Our Doppler-imaging code \emph{iMap} \citep{2012A&A...548A..95C} performs a temperature inversion for a number of photospheric line profiles simultaneously. For the inversions 20 suitable absorption lines were selected from the 5000--6750\,\AA\ wavelength range \citep[for the selection criteria see][]{2015A&A...578A.101K}. Each spectral line is modeled individually and locally, then being disk-integrated and in the final step all disk-integrated line regions are averaged to obtain a mean theoretical profile. These mean profiles are then compared with each observed mean profile \citep[for more details see Sect. 3 in][]{2012A&A...548A..95C}. For the preparation of the observed mean profile
we proceed with a simple S/N-weighted averaging to increase the overall S/N by a factor of $\approx$4. \emph{iMap} calculates the local line profiles by solving the radiative transfer with the help of an artificial neural network \citep{2008A&A...488..781C}. Atomic parameters are taken from the VALD database \citep{1999A&AS..138..119K}. Model atmospheres are taken from \citet{2004astro.ph..5087C} and are interpolated for each desired temperature, gravity and metallicity. Due to the high CPU demand only LTE radiative transfer is used instead of spherical non-LTE model atmospheres.

For the  surface reconstruction \emph{iMap} uses an iterative regularization based on a Landweber algorithm \citep{2012A&A...548A..95C}. According to our tests \citep[see Appendix A in][]{2012A&A...548A..95C} the iterative regularization has been proven to converge always on the same image solution. Therefore, no additional constraints are imposed for the image reconstruction. The surface element resolution is set to $5^{\circ} \times 5^{\circ}$.

\begin{figure*}[]
\vspace{2.0cm}{\hspace{0.5cm}\Large{S01}}

\vspace{0.3cm}{\hspace{0.4cm}\large{Jan-30}}
\vspace{-2.9cm}

\hspace{2.00cm}\includegraphics[angle=0,width=1.74\columnwidth]{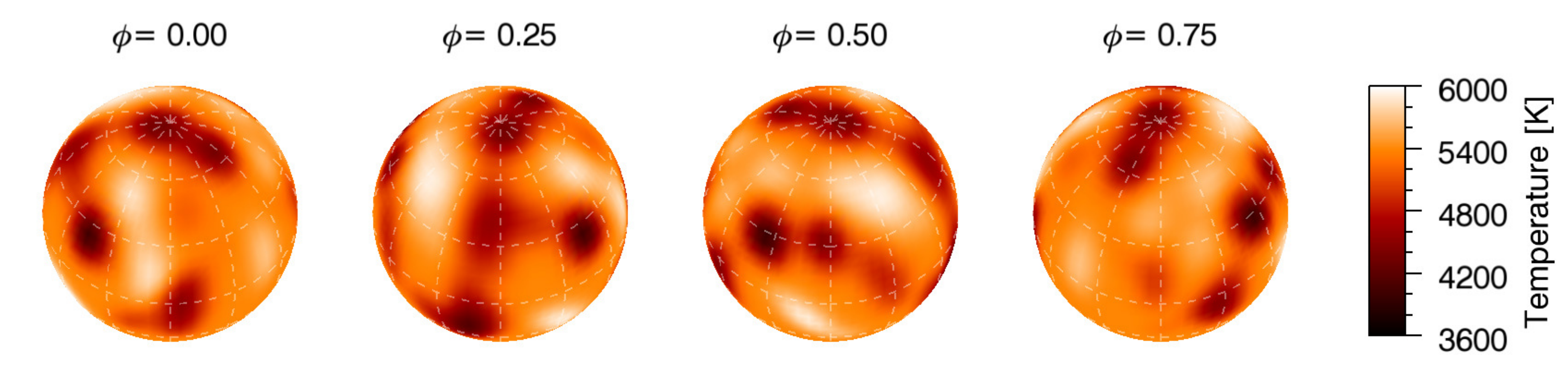}

\vspace{2.0cm}{\hspace{0.5cm}\Large{S02}}

\vspace{0.3cm}{\hspace{0.4cm}\large{Feb-07}}
\vspace{-2.9cm}

\hspace{2.00cm}\includegraphics[angle=0,width=1.74\columnwidth]{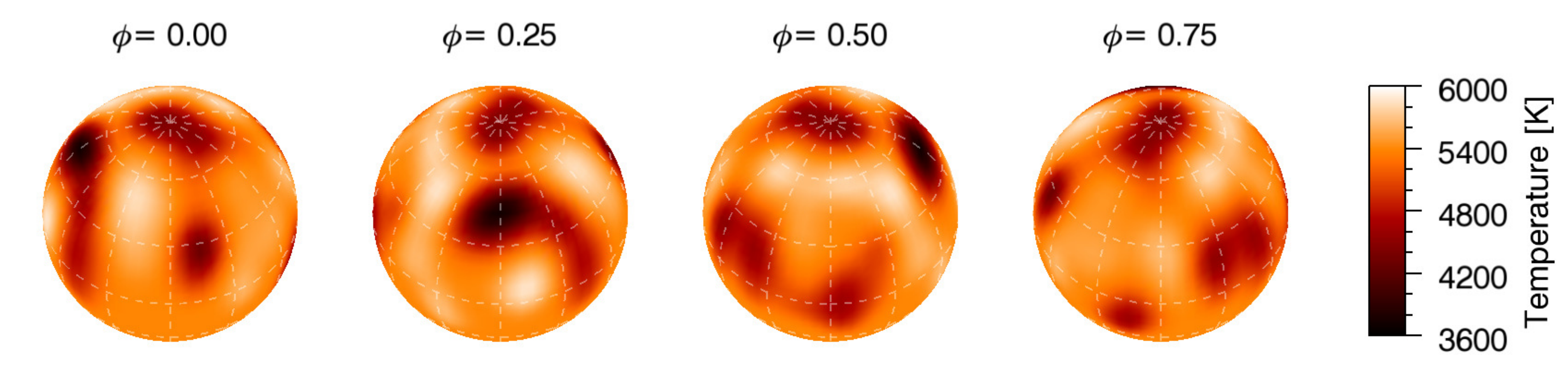}

\vspace{2.0cm}{\hspace{0.5cm}\Large{S03}}

\vspace{0.3cm}{\hspace{0.4cm}\large{Feb-28}}
\vspace{-2.9cm}

\hspace{2.00cm}\includegraphics[angle=0,width=1.74\columnwidth]{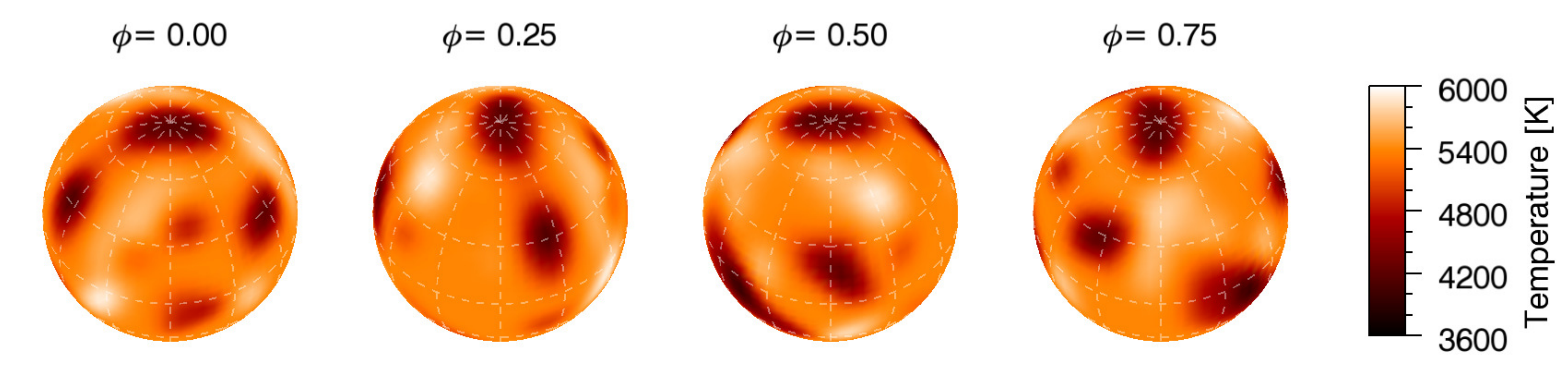}

\vspace{2.0cm}{\hspace{0.5cm}\Large{S04}}

\vspace{0.3cm}{\hspace{0.4cm}\large{Mar-07}}
\vspace{-2.9cm}

\hspace{2.00cm}\includegraphics[angle=0,width=1.74\columnwidth]{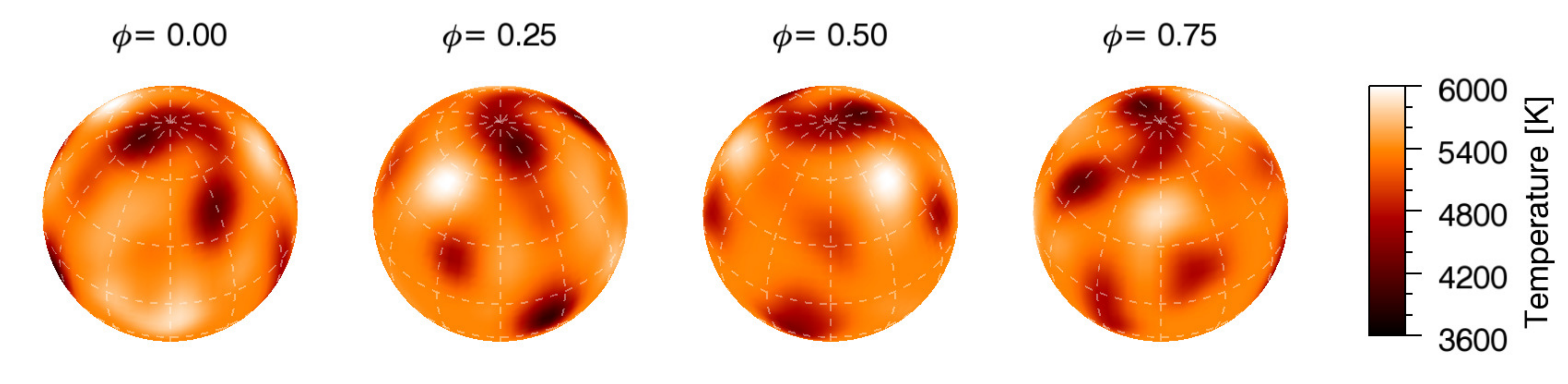}

\vspace{2.0cm}{\hspace{0.5cm}\Large{S05}}

\vspace{0.3cm}{\hspace{0.4cm}\large{Mar-13}}
\vspace{-2.9cm}

\hspace{2.00cm}\includegraphics[angle=0,width=1.74\columnwidth]{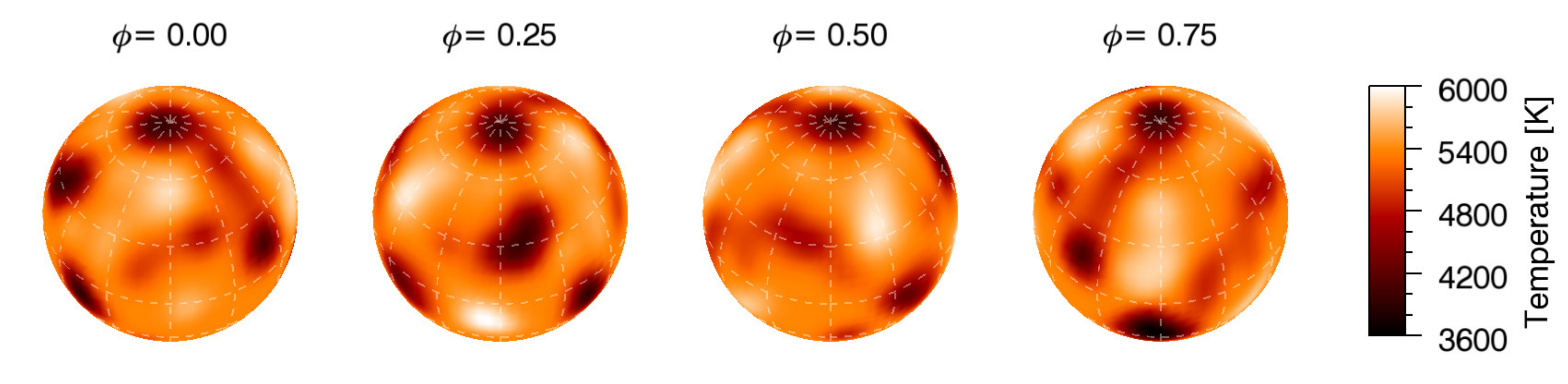}

\vspace{2.0cm}{\hspace{0.5cm}\Large{S06}}

\vspace{0.3cm}{\hspace{0.4cm}\large{Mar-23}}
\vspace{-2.9cm}

\hspace{2.00cm}\includegraphics[angle=0,width=1.74\columnwidth]{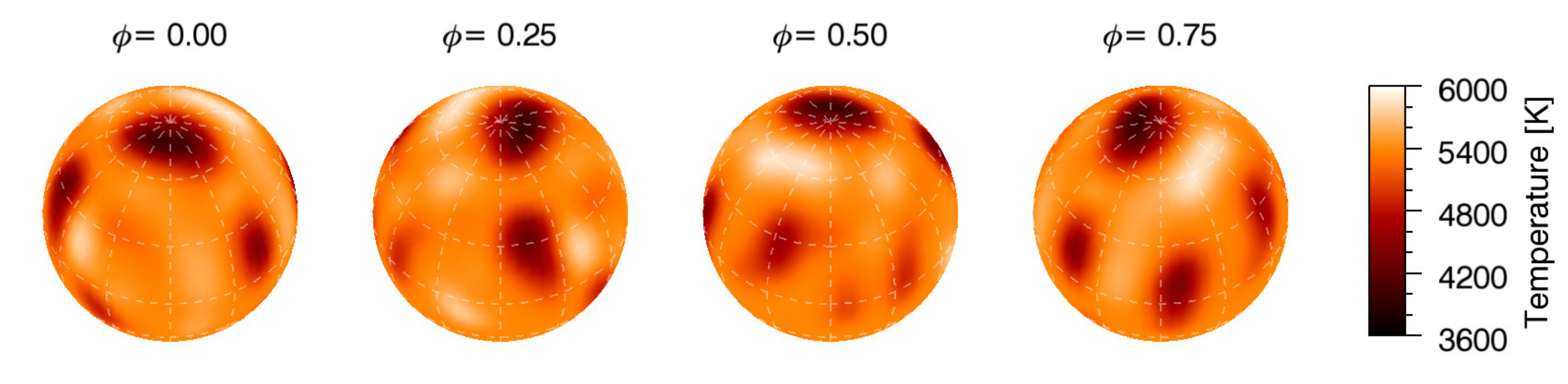}

\caption{Doppler images of \inc for STELLA data sets S01--S06. The corresponding mid-UT dates (2017+) are indicated below the names of the maps.}
\label{dis1}
\end{figure*}

\begin{figure*}[]
\vspace{2.0cm}{\hspace{0.5cm}\Large{S07}}

\vspace{0.3cm}{\hspace{0.4cm}\large{Mar-28}}
\vspace{-2.9cm}

\hspace{2.00cm}\includegraphics[angle=0,width=1.74\columnwidth]{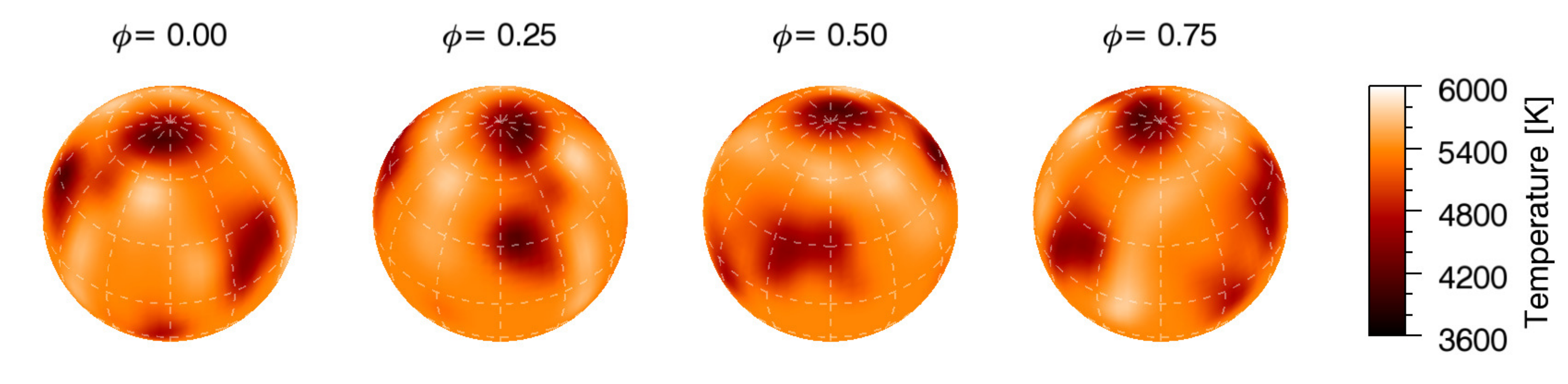}

\vspace{2.0cm}{\hspace{0.5cm}\Large{S08}}

\vspace{0.3cm}{\hspace{0.4cm}\large{Apr-20}}
\vspace{-2.9cm}

\hspace{2.00cm}\includegraphics[angle=0,width=1.74\columnwidth]{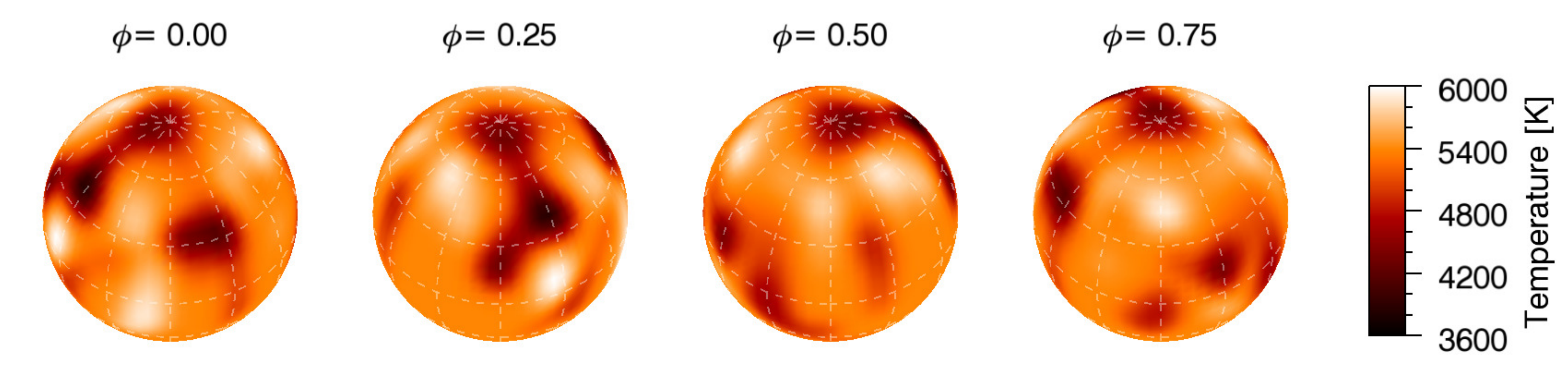}

\vspace{2.0cm}{\hspace{0.5cm}\Large{S09}}

\vspace{0.3cm}{\hspace{0.4cm}\large{May-03}}
\vspace{-2.9cm}

\hspace{2.00cm}\includegraphics[angle=0,width=1.74\columnwidth]{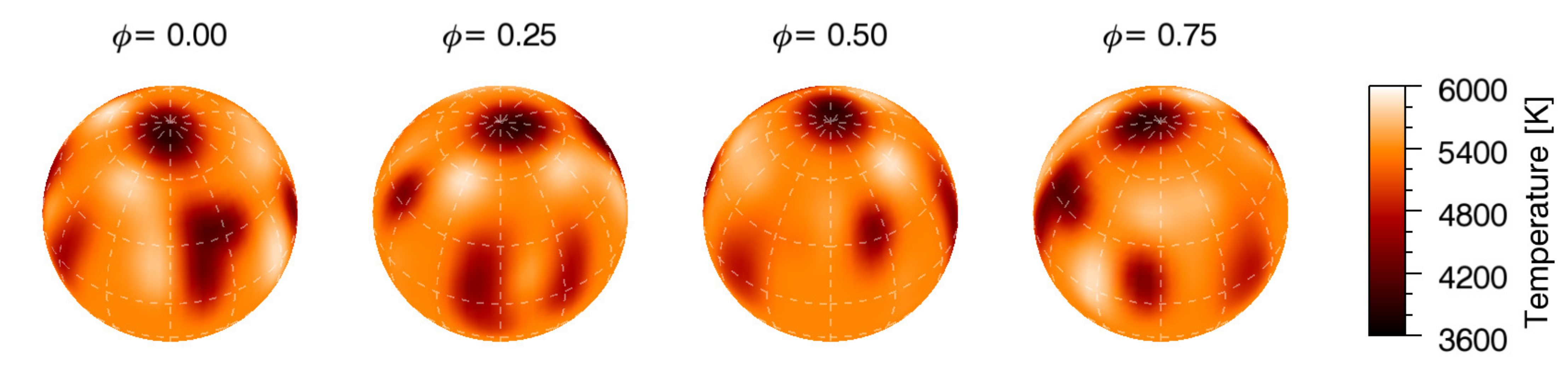}

\vspace{2.0cm}{\hspace{0.5cm}\Large{S10}}

\vspace{0.3cm}{\hspace{0.4cm}\large{May-19}}
\vspace{-2.9cm}

\hspace{2.00cm}\includegraphics[angle=0,width=1.74\columnwidth]{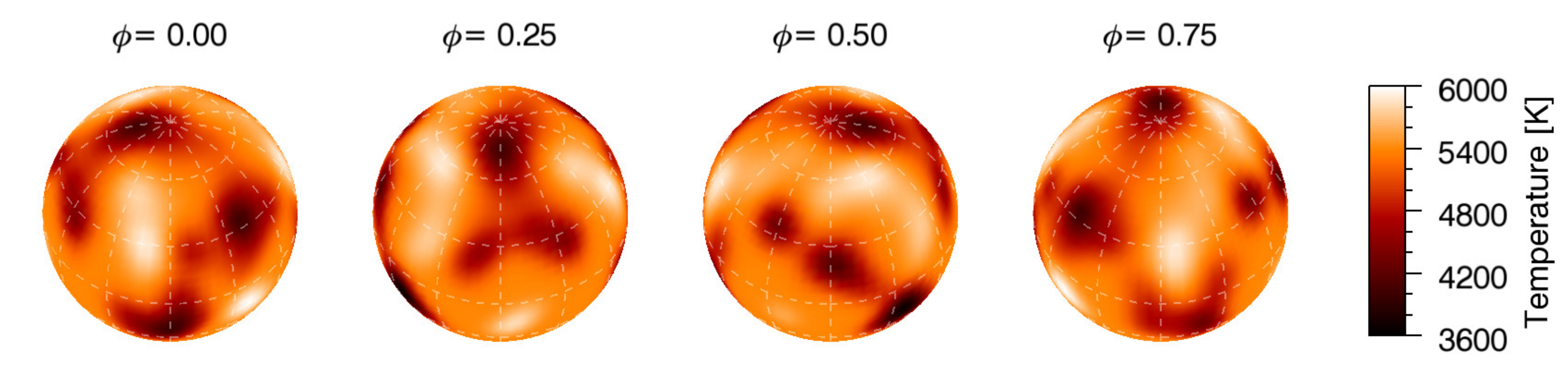}

\vspace{2.0cm}{\hspace{0.5cm}\Large{S11}}

\vspace{0.3cm}{\hspace{0.4cm}\large{May-25}}
\vspace{-2.9cm}

\hspace{2.00cm}\includegraphics[angle=0,width=1.74\columnwidth]{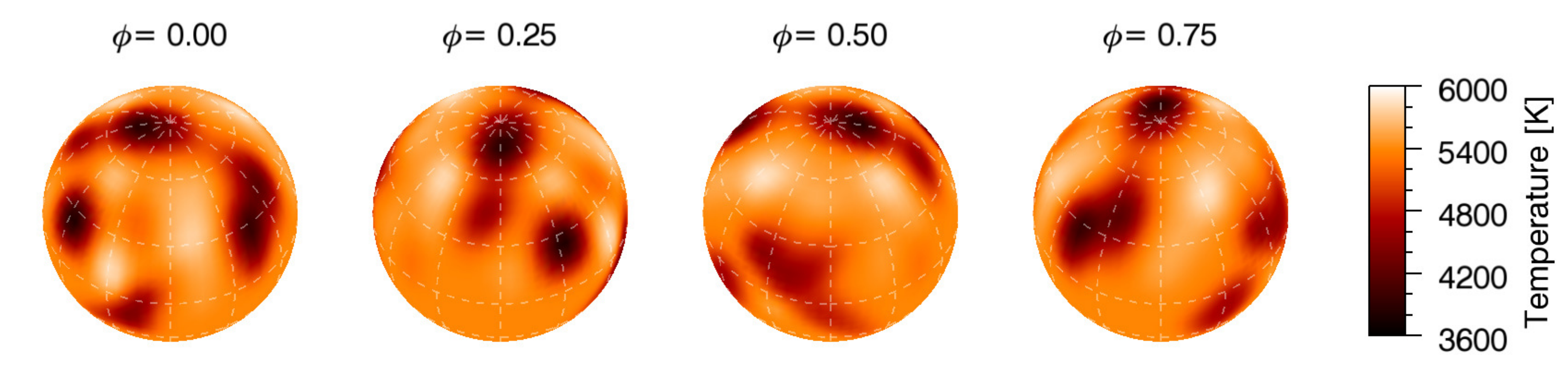}

\vspace{2.0cm}{\hspace{0.5cm}\Large{S12}}

\vspace{0.3cm}{\hspace{0.4cm}\large{Jun-08}}
\vspace{-2.9cm}

\hspace{2.00cm}\includegraphics[angle=0,width=1.74\columnwidth]{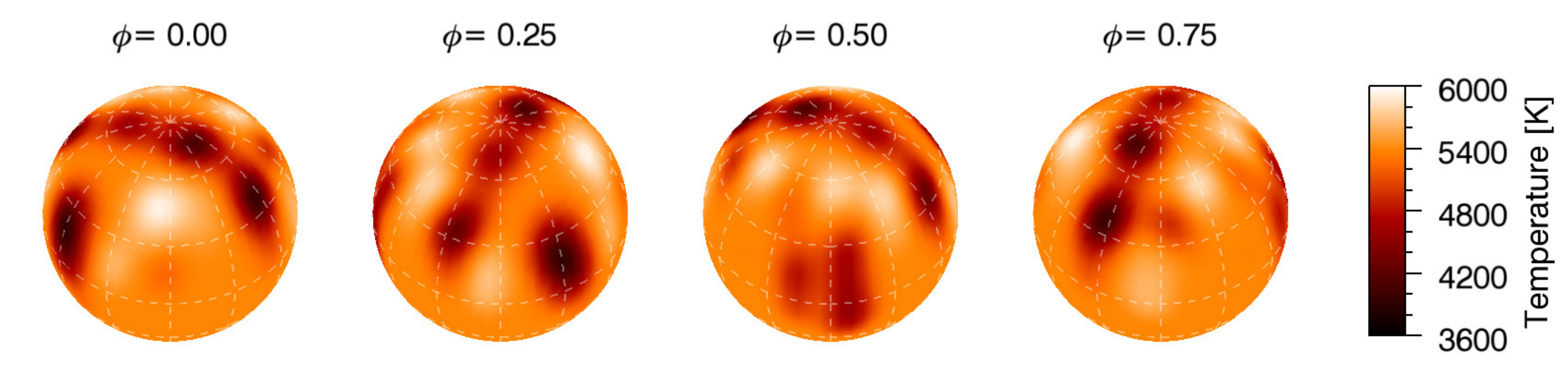}

\caption{Doppler images for STELLA data sets S07--S12. The corresponding mid-UT dates (2017+) are indicated below the names of the maps.}
\label{dis2}
\end{figure*}

\begin{figure*}[th]
\vspace{2.0cm}{\hspace{0.5cm}\Large{P01}}

\vspace{0.3cm}{\hspace{0.4cm}\large{Mar-11}}
\vspace{-2.9cm}

\hspace{2.00cm}\includegraphics[angle=0,width=1.74\columnwidth]{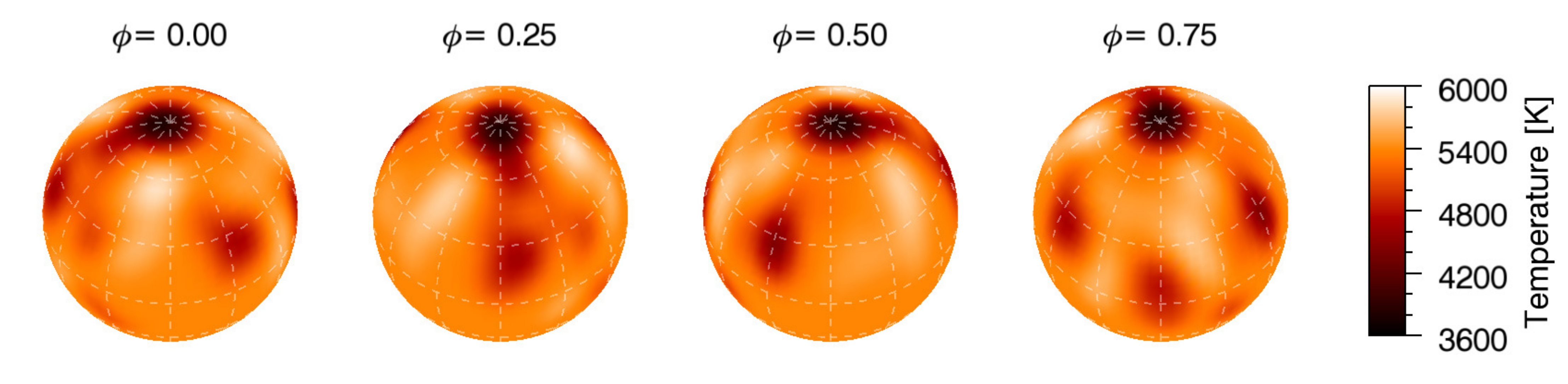}

\caption{Doppler image for the PEPSI@VATT spectra. The corresponding mid-UT date is 2017-03-11, which falls just between the dates of S04 and S05 maps shown in Fig.~\ref{dis1}.}
\label{dipepsi}
\end{figure*}

\subsection{Results: spot morphology and evolution}

The resulting 13 time-series Doppler reconstructions of \inc for 2017 (12 maps for the STELLA observations and one for the PEPSI spectra) are plotted in Figs.~\ref{dis1}, \ref{dis2} and \ref{dipepsi}. The line-profile fits are given in the Appendix in Figs.~\ref{proffits1}, \ref{proffits2} and \ref{proffits3}, respectively. The overall surface structure characteristics is reminiscent of the first and so far only Doppler image from 1994 (see Paper~I), revealing a cool spot on the visible pole, cool spots at low to high latitudes and even a few hot spots mostly at mid-latitudes. The spot temperatures range from the coolest $\approx$3600\,K up to $\approx$5800\,K, i.e., hotter by 400\,K than the unspotted photosphere of 5400\,K. Hot spots are often claimed to be artifact of the imperfect reconstruction. \citet[][see their Fig. 3]{2014A&A...562A.139L} have demonstrated that such artifacts are usually the result of extremely sparse phase coverage. However, regarding the reconstructions as a time series we observe that subsequent maps reveal quite similar features (cool as well as hot spots) from totally independent data. Moreover, insufficient phase sampling would introduce artificial (hot as well as cool) features at different locations from one Doppler reconstruction to the next, i.e., usually around phases where the largest phase gaps appear. However, our datasets are well sampled, and their largest gaps (usually below 0.15-0.17 phase fraction, i.e. still not very large) appear randomly along the rotation phase, therefore we do not expect such artificial hot (and cool) features at similar locations over the 13 individual Doppler reconstructions. Note especially the P01 PEPSI-map shown in Fig.~\ref{dipepsi}, which falls between the S04 and S05 STELLA-maps and despite the different observing facilities and independent data the recovered surface features show remarkable resemblance. This confirms not only the reliability of the reconstructed features but also the steadiness and robustness of \emph{iMap}. The polar spot seems to be the most permanent feature over the time range, while at lower latitudes the spotted surface is more variable, still, the dominant features can be tracked from one map to the next. Finally, we note that in some maps strong features are seen also below the equator, despite that Doppler imaging is less powerful when reconstructing the less visible hemisphere. We assume, however, that such a feature is most likely real when it reappears on consecutive Doppler reconstructions (see S03-S04-S05 and S10-S11), although the shape, size or contrast of these features may be loose.

The first reconstruction in the time series (S01) reveals an elongated polar feature of $\approx$4800\,K together with several lower latitude nearly circular spots of $\approx$4000$-$5000\,K with typically $10^{\circ}$ diameter. The brightest feature of $\approx$5800\,K is centered at phase $\phi$$=$0.4 at high ($\approx$50$^{\circ}$) latitude. A faint cool spot at $\phi$$=$0.25 is becoming the most prominent cool feature for the next map (S02), while the hot spot as well as the other cool spots are getting less contrasted.  For the next map (S03) the polar spot is getting cooler and more compact, while other cool spots are shrinking by $\approx$30-60\%. The only exception is the new cool feature at the lower hemisphere, just at the border of visibility. We note that the bright spot at $\phi$$=$0.4 is permanently visible. For the next (S04) map the cool spots are becoming fainter, however, the bright spot at $\phi$$=$0.4 is hotter. S05, the fifth map reveals an emerging new cool spot at $\phi$$\approx$0.3, while the polar spot has become more compact and contrasted. In the next map (S06) the progeny of the new spot, as well as the other cool and bright features, become smaller and/or less contrasted. This continues in S07 map, where the hot features nearly vanish. Traces of new flux emergence are seen in S08 with a new spot at $\phi$$\approx$0.2. Also, bright features appear again, in particular the well-known one at $\phi$$=$0.4. In the next map (S09) the polar spot is getting more prominent while the newly emerged spot at $\phi$$\approx$0.2 fades and splits into two subspots. The high latitude bright spot at $\phi$$=$0.4 is still detectible. In the tenth map (S10) a new cool spot group emerged at $\phi$$\approx$0.5, and is getting less contrasted and shifted towards the covered pole in the S11 map. Also, the formerly vanishing spot around 0.2 phase appears now strengthened. This continues during our last reconstruction (S12), where the polar spot is shrinking and also displacing. Permanent rearrangements are taking place, e.g. the spot group in S11 at $\phi$$\approx$0.9 is getting smaller in size and cooler. Note also, that the high latitude warm ($\approx$5600$-$5900\,K) features are still present.

\begin{figure}[]
\includegraphics[angle=0,width=1\columnwidth]{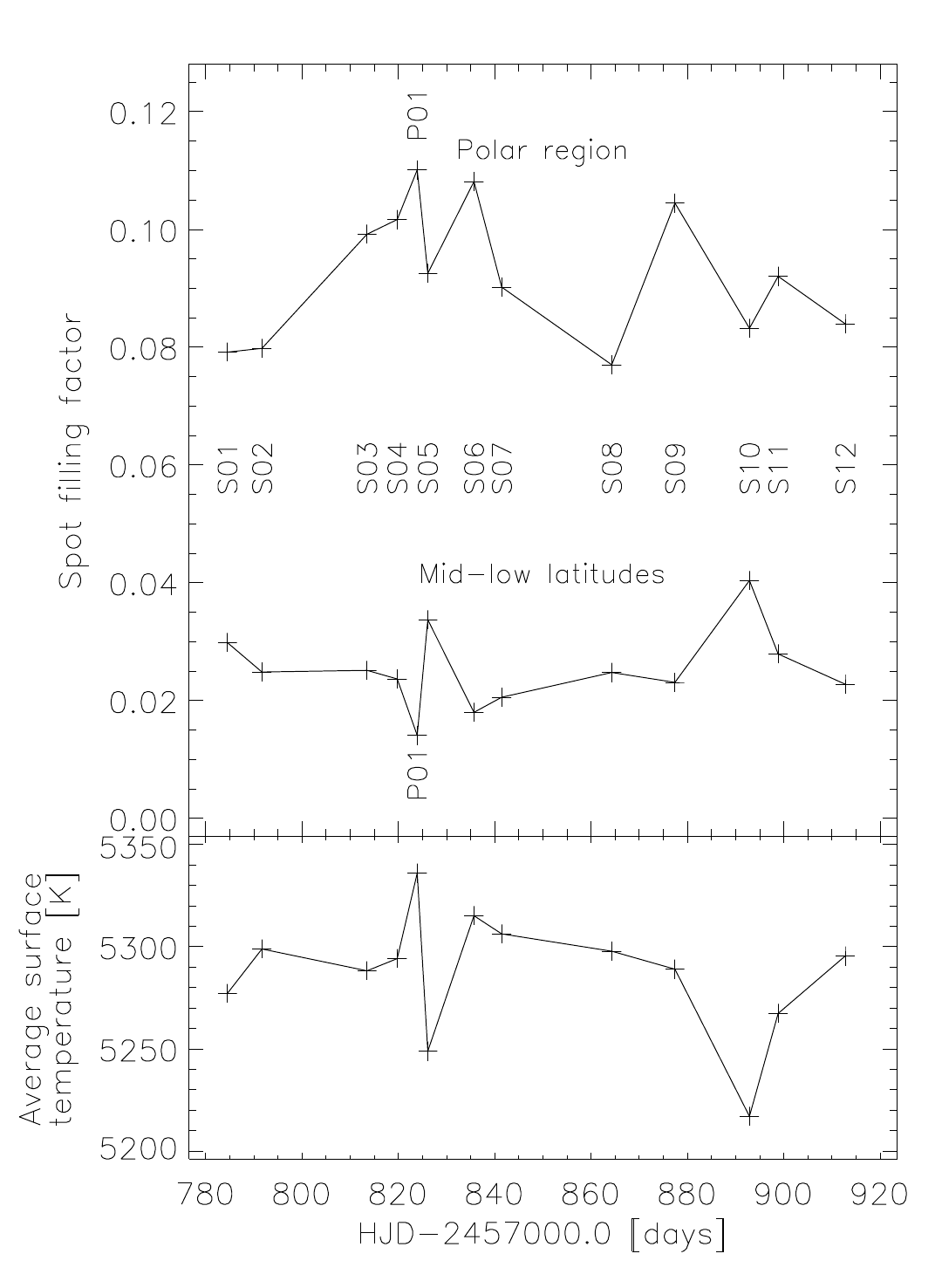}
\caption{Time variation of the spot filling factor (top panel) and the integrated surface temperature (bottom panel) of \inc derived from the time-sereies Doppler images shown in Figs.~\ref{dis1}, \ref{dis2} and \ref{dipepsi}. In the upper panel the spot filling factors are shown for the polar region and for the mid-to-low latitudes separately; see the top and the bottom curves, respectively. The different Doppler reconstructions are identified by their serial numbers.}
\label{spotsvstime}
\end{figure}

For each map the overall surface temperature is obtained by averaging the temperature values pixel by pixel over the stellar surface. However, tracing individual spots from one map to the next is hampered by the rapid spot rearrangements and/or the imperfect phase coverages. Instead, we split the surface into two parts above and below 65\degr, this way ranking the spots to be either polar or low-to-mid latitude spots. We measure the time variation of both surface partitions by deriving the spot filling factor values. In Fig.~\ref{spotsvstime} we plot the time variation of the average temperature from surface integration as well as the spot filling factors. The diagrams indicate two epochs at HJD~2\,457\,826 (S05) and HJD~2\,457\,893 (S10), when the average temperature decreased by $\approx$50\,K, simultaneously with a small drop of the filling factor at the pole, but a significant increase of $\approx$40\% at mid-low latitudes. According to the maps, these two events may indicate significant spot rearrangements, when new fluxes emerge. On the other hand, individual spot evolutions imply that the average spot lifetime should be of the order of a month.

\subsection{Surface differential rotation}\label{ccf}

Tracking short term spot migrations is among the usual methods to study stellar surface differential rotation from Doppler imaging \citep{1997MNRAS.291....1D}. In this paper, we apply the program \texttt{ACCORD} \citep[][and references therein]{2015A&A...573A..98K} and perform a time-series cross-correlation analysis from the 12 Doppler images obtained for the STELLA observations (for the sake of data homogeneity we excluded the P01 map from this analysis). It provides 11 consecutive cross-correlation function (ccf) maps which are combined into an average correlation map.  Its 2D correlation pattern is then fitted with a quadratic differential-rotation law in the usual (solar) form of $\Omega(\beta)=\Omega_{\rm eq}(1-\alpha\sin^2\beta)$, where $\Omega(\beta)$ is the angular velocity at latitude $\beta$, $\Omega_{\rm eq}$ the angular velocity at the equator, while $\alpha=(\Omega_{\rm eq}-\Omega_{\rm pole})/\Omega_{\rm eq}$ is the relative angular velocity difference between the equator and the pole, i.e. the surface shear coefficient.

\begin{figure}[]
\includegraphics[width=1.0\columnwidth]{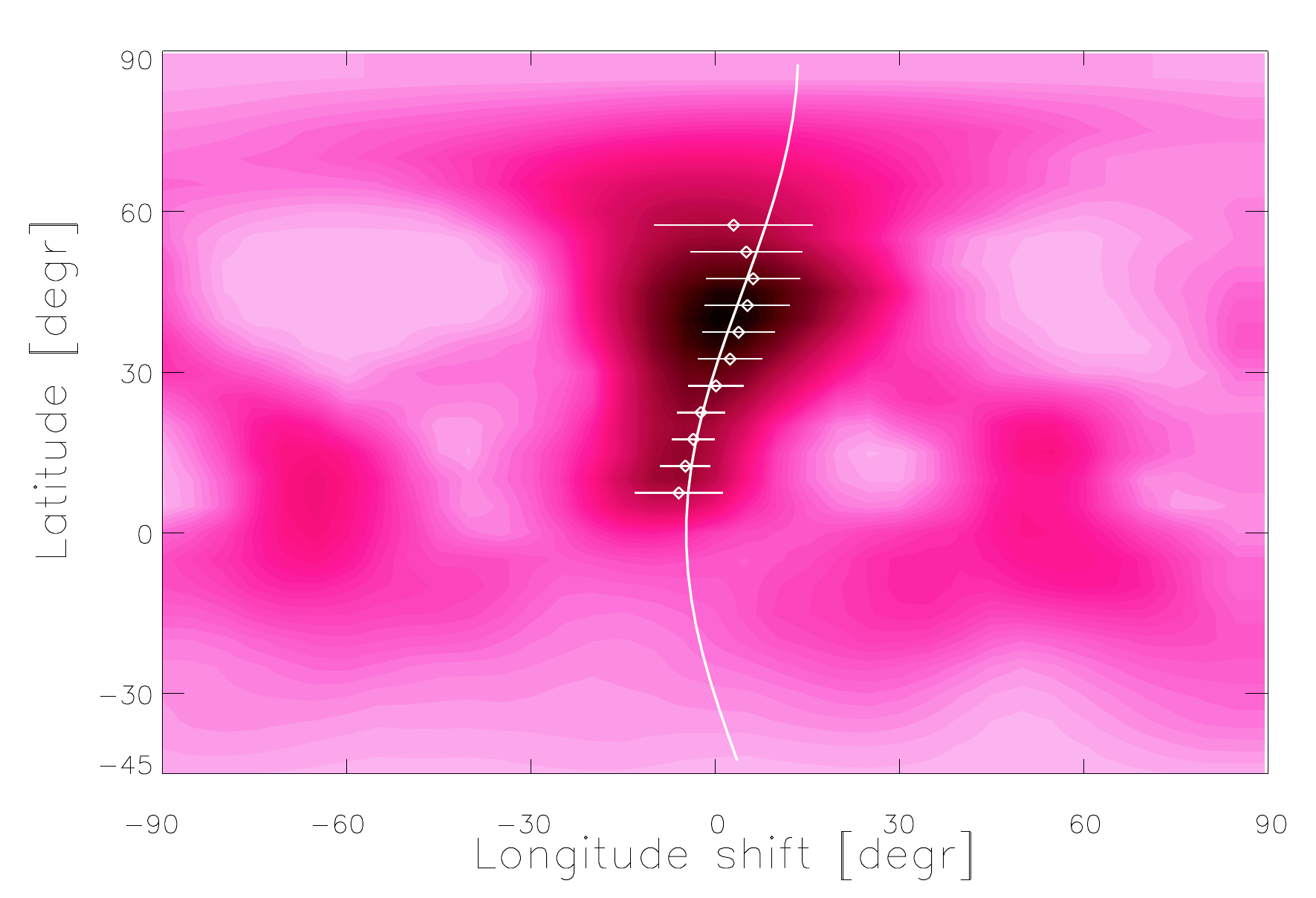}
\caption{Average cross-correlation map for \inc showing anti-solar surface differential rotation.  The best correlated dark regions are fitted by Gaussian curves in 5\degr\ bins. Gaussian peaks are indicated by dots, the corresponding Gaussian widths by horizontal lines. The best fit differential rotation law suggests an equatorial period of $P_{\rm eq}=5.973$\,d and a surface shear coefficient of $\alpha=-0.026$.}
\label{ccf_iMap}
\end{figure}

The resulting correlation pattern for \inc is shown in Fig.~\ref{ccf_iMap}. It indicates anti-solar surface differential rotation, i.e., the equator rotates slower than the polar latitudes. The most well correlated dark regions are fitted with Gaussian curves in 5\degr\ bins. The Gaussian peaks are indicated in Fig.~\ref{ccf_iMap}. The best fit to these peaks gave $\Omega_{\rm eq}=60.28\pm0.08$\,\degr/d or an equivalent equatorial period of $P_{\rm eq}=5.973\pm0.008$\,d with a shear coefficient of $\alpha=-0.026\pm0.005$.  This yields a lap time of 230\,d needed by the polar regions to lap the equator by one full rotation. 

\section{Variability of the IN Comae system}\label{phot}

\subsection{Orbital photometric modulation}

\begin{figure}[]
\includegraphics[angle=0,width=1\columnwidth]{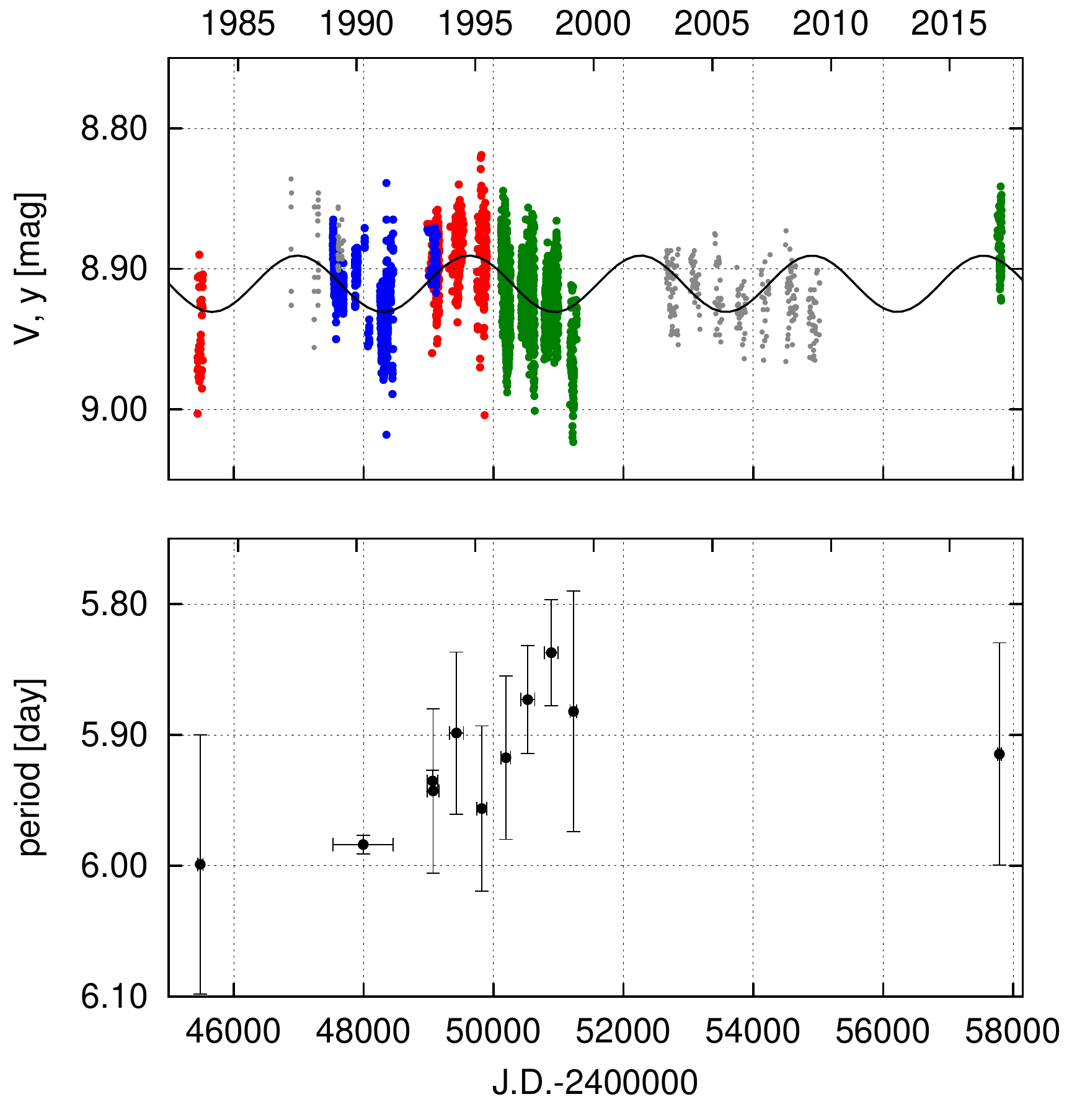}
\caption{Top: long-term photometric $V$ and $y$ data of \ind. Different colors mean different sources of observations; green: data from T6 and T7 APTs, blue: observations from the Hungarian 1-m RCC telescope, red: literary data (mostly from Paper~1) which are used for seasonal period determination, grey: literary data+ASAS data which are not suitable for seasonal period determination. The sine-wave fit by black solid line represents the long-term overall brightness change with a period of 7.2 years, i.e., basically the wide binary period, see Sect.~\ref{disc}. Bottom: independent rotational period determinations for suitable seasonal datasets. See text for details.}\label{figphot}
\end{figure}

Fig.~\ref{figphot} presents photometric data of \inc for the past 30+ years. To support a long-period search our new photometric data are combined with the published observations from Paper~I and augmented with observations from the All Sky Automated Survey (ASAS) database \citep{2002AcA....52..397P}. The (binned) SuperWASP data in Aller et al. (2018) could not be used due to missing bandpass transformations but overlap with part of the ASAS data anyway. For the period determination, we apply the Fourier-transformation based frequency analyzer code MuFrAn \citep{2004ESASP.559..396C}. In the top panel of Fig.~\ref{figphot}, we show the best fit to the full photometric data set with a sinusoid of a period of 2639\,d ($\approx$7.2\,yr), which has an uncertainty of about 200\,d. The sine-wave fits well the first two well-observed cycles and does not contradict with the later, sparse data.

\begin{figure}[]
\includegraphics[width=1\columnwidth]{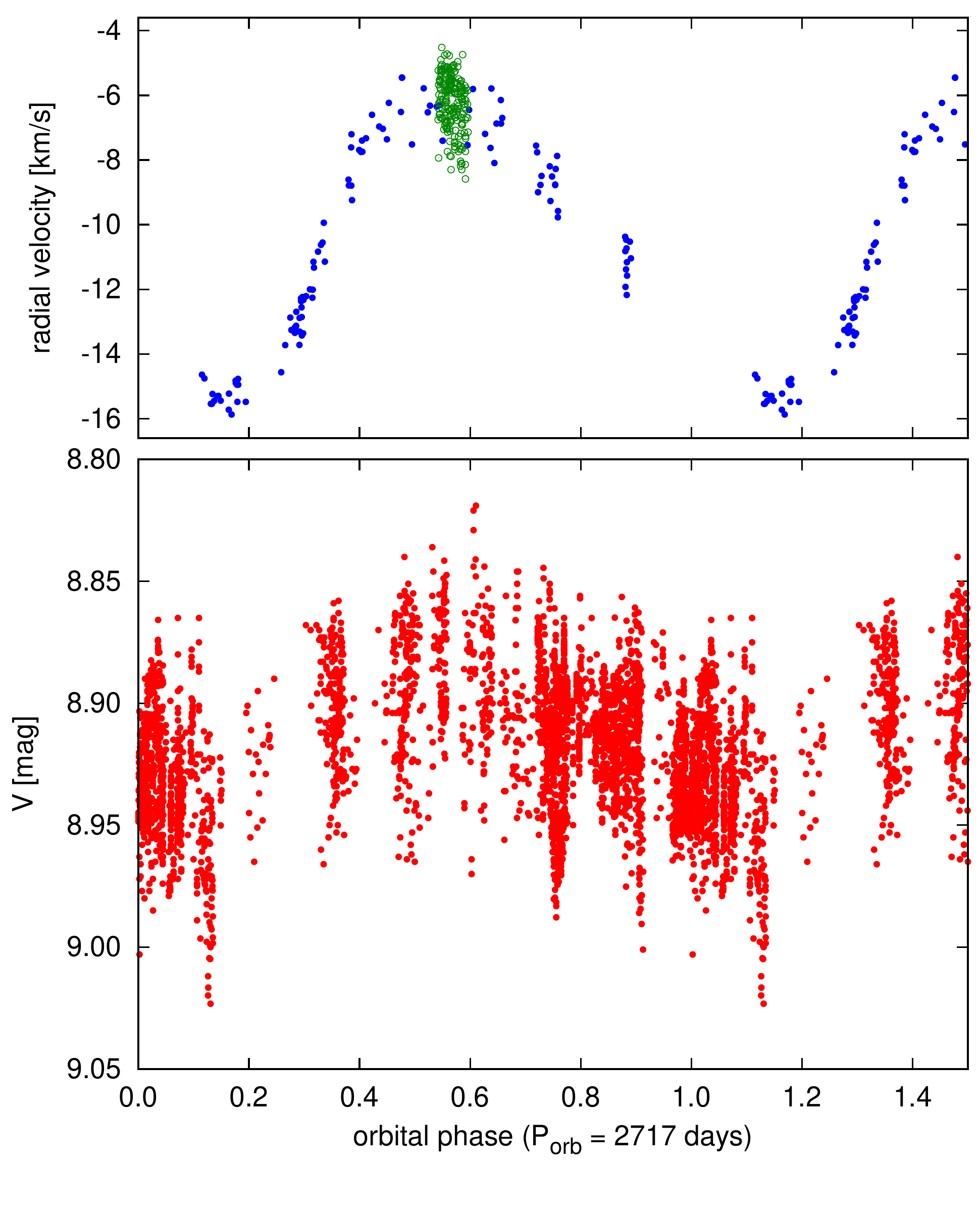}
\caption{Comparing the long-term light variation with the radial velocity curve of \ind. Top: radial velocity curve of the star taken from \citet{2017A&A...600L...9J}; suggesting a 2717-day long orbital period. Overplotted are the radial velocities from our spectroscopic data (green circles). Bottom: the long-term $V+y$ photometric observations after folding up with the orbital period.}\label{lcrad}
\end{figure}

The photometric cycle of 2639\,d is in surprising agreement with the recently proposed orbital period of 2717$\pm$63\,d \citep{2017A&A...600L...9J} (and also with its revised value of 2689$\pm$52\,d by Aller et al. 2018). Orbital phase coherence of surface activity is common in comparably short period tidally-connected RS\,CVn binaries, but has never been seen for such long period timescales. Fig.~\ref{lcrad} shows a comparison of the orbital radial velocities with our long-term APT photometry phased with the same orbital period of 2717 days from Jones et al. (2017). Most notable is the phase coherence in the sense that the light-curve maximum coincides with a time of highest radial velocity while the light-curve minimum coincides with a time of lowest radial velocity. 

\subsection{Rotational photometric modulation}\label{rotmod}

Seasonal short period determinations are presented in the lower panel of Fig.~\ref{figphot}. We note that the amplitude of the rotational modulation of \inc is generally low, typically less than 0\fm1 in $V$. Therefore, seasonal rotational periods were derived only for the best quality datasets with good phase coverage and low scatter. The average value of the seasonal periods is $\approx$5.92 days. At this point we emphasize that any photometric period always traces the rotation period of the star at that latitude where the spot or spots occurred. Interestingly, between 1995--1999, the photometric  period was increasing, while the overall brightness was decreasing. Such a simultaneity is explained by surface differential rotation that causes a shift of the dominant longitude usually populated by star spots \citep[cf.][]{2014MNRAS.441.2744V}. 

\begin{figure}[hb]
\includegraphics[angle=0,width=1\columnwidth]{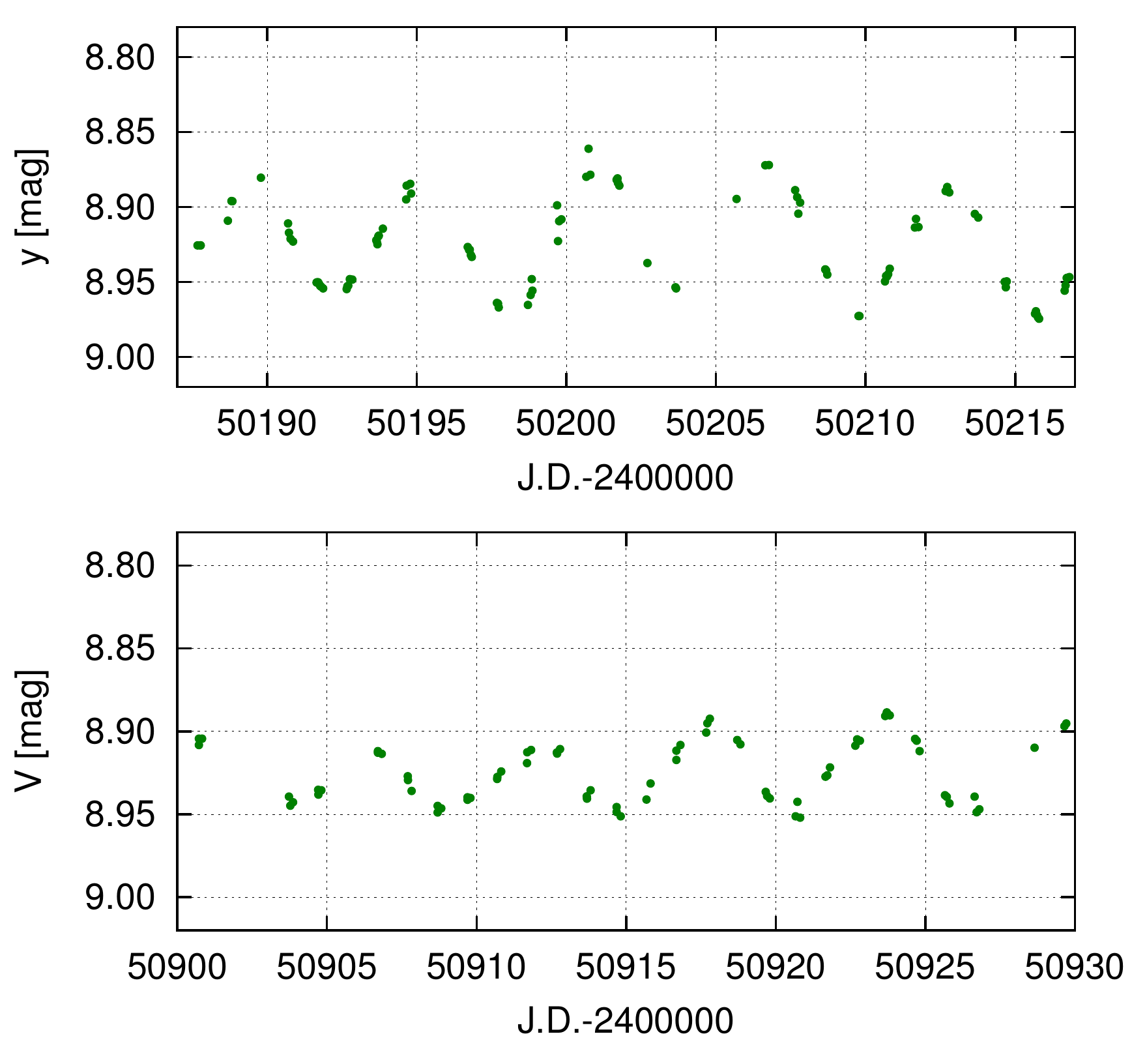}
\caption{Well sampled photometric modulation indicates the solidity of the 5.9-day period. Light curves were taken during April/May 1996 (top) and April 1998 (bottom) with the T7 and T6 APTs, respectively. The light curves even show the changing amplitudes and shapes typical for star spot evolution.}
\label{2lc}
\end{figure}

Besides, our data again demonstrate the solidity of the 5.9-day photometric period being the rotation period as opposed to, e.g., the 1.2-day ($1-f$) alias. This was already done in our Paper~I but then we had not had a beautiful photometric light curve that sampled the variation with high-enough time resolution. Here we present two independent, densely sampled and time-continuous APT data (Fig.~\ref{2lc}) that proof without doubt that the 5.9-day period is indeed the correct one.

\begin{figure}[thb]
\includegraphics[angle=0,width=1\columnwidth]{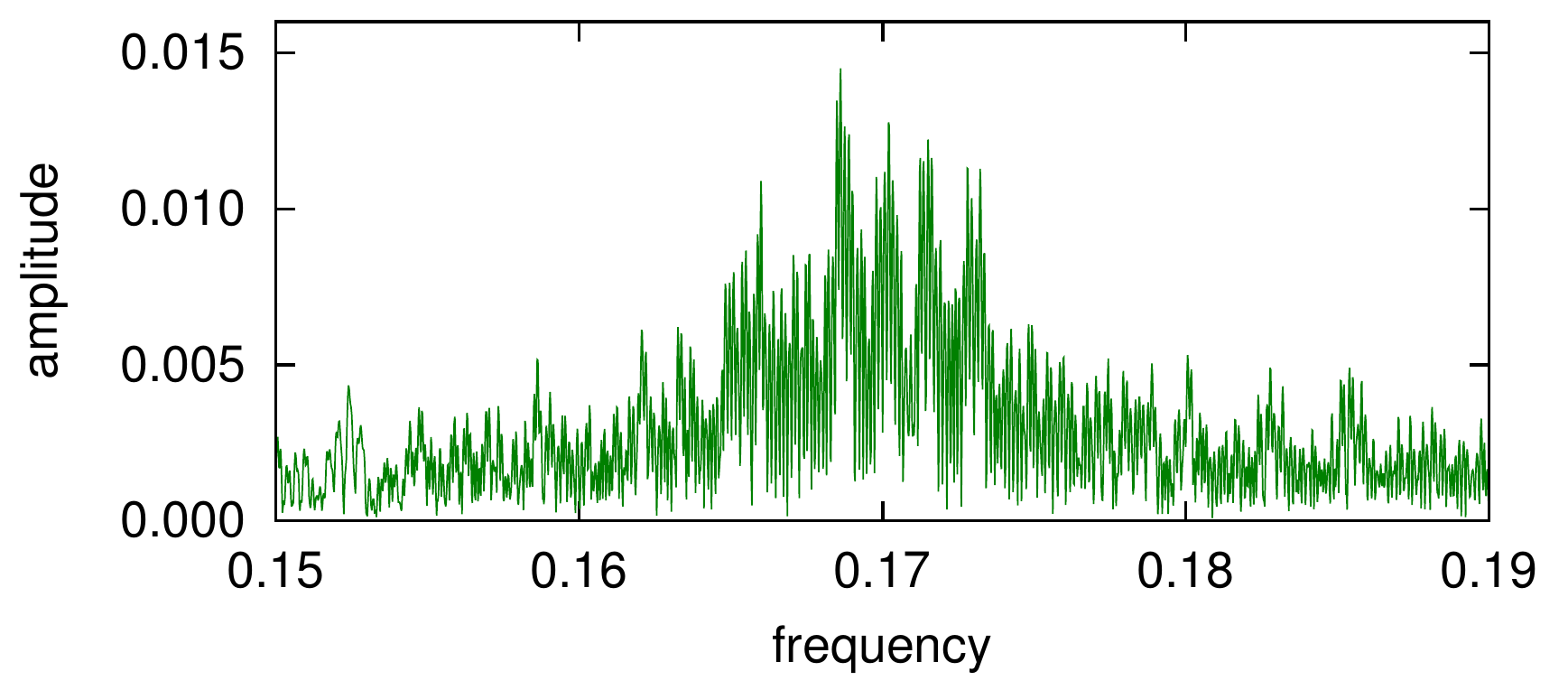}

\includegraphics[angle=0,width=1\columnwidth]{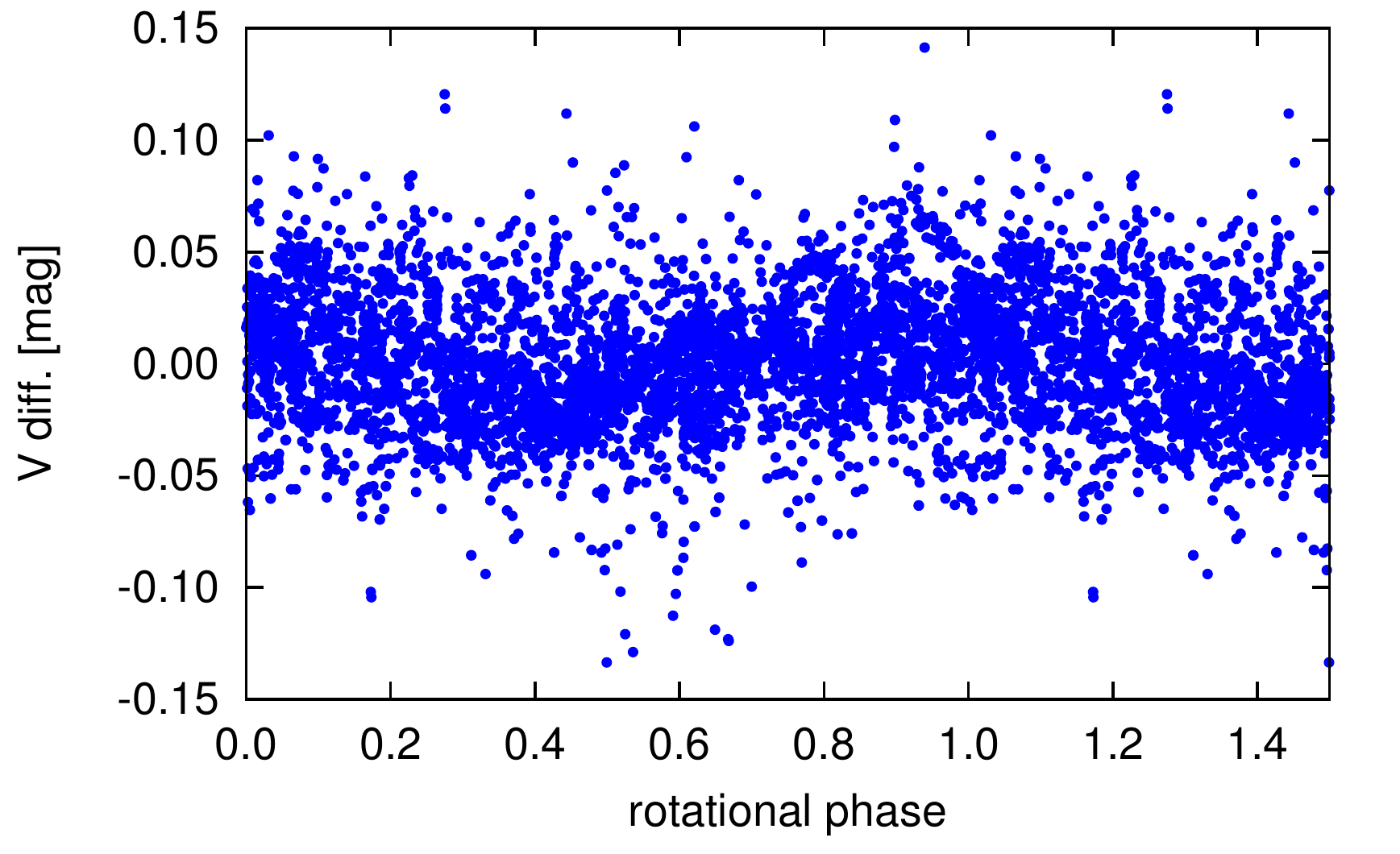}
\caption{Top: Fourier-amplitude spectrum for all the combined $V$ and $y$ photometric data shown in Fig.~\ref{figphot}. Bottom: $V$ and $y$ data folded with $P_{\rm rot}=5.934$ days.}
\label{Fourier}
\end{figure}

The next step is to average the individual photometric period by applying a period search to the full data set. For this, we combine all the available high-precision $V$ and $y$ data into one single dataset and analyze it with the time-frequency analysis package MuFrAn. The resulting Fourier-amplitude spectrum is obtained for the pre-whitened data, that is the data with the long-term trend and the 2849-d period removed. The resulting amplitude spectrum is shown in the top panel of Fig.~\ref{Fourier}. Its highest peak suggests a long-term average photometric period of 5.934$\pm$0.001\,d, which is very close to the average of the seasonal values. Consistently with the result of the differential rotation analysis in Sect.~\ref{ccf_iMap} this period is the apparent rotation period of the mid-latitude belt around $\approx$35\degr where spots cause the most significant light variation. Accordingly, in the bottom panel of Fig.~\ref{Fourier}, we fold the 30+ years of photometry with this period. It shows a  phase coherency that is remarkable over that period of time. Thus, for future phase calculations, we suggest to use the following equation 
\begin{equation}\label{eq1}
{\rm HJD} = 2,449,415.0 + 5.934\times E ,
\end{equation}
where the reference time was taken from Paper~I. This is also the ephemeris that we used to phased our Doppler images in Sect.~\ref{di}.

Another finding is the frequency splitting of the dominant Fourier peak at 0.17\,d$^{-1}$. The surrounding lower amplitude peaks are typical signature of differential surface rotation of the star  \citep[see the simulations in][]{2004ESASP.538..149S}. Its individual peaks mark the stellar latitudes where the spots preferentially occurred. The five most prominent peaks of the Fourier-spectrum are listed in Table~\ref{Fpeaks}. Assuming that these peaks are due to surface differential rotation and the lowest and highest frequencies correspond to the polar and equatorial regions (or vice versa), we estimate a surface shear parameter of $\Delta P/P\approx0.03$, albeit without any presumption on its sign. Such an estimation is usually erroneous, since the origin of the photometric signals is ambiguous (e.g., it is not known at which stellar latitudes the signaling spots are located). Nevertheless, this value is of the same order as we found from the cross-correlation analysis in Sect.~\ref{ccf} and confirms the existence of strong differential rotation on this G5 giant.

\begin{table}[b]
\caption{The five most prominent periods and their wave amplitudes from the light curve Fourier-analysis.}\label{Fpeaks}
 \centering
 \begin{tabular}{ccc}
  \hline
  \hline\noalign{\smallskip}
Frequency & Amplitude &   Period \\
(1/d)     & (mag)     &   (d) \\
   \hline
  \noalign{\smallskip}
 0.168506  & 0.0143 & 5.934 \\
 0.171509  & 0.0130 & 5.831 \\
 0.172930  & 0.0130 & 5.783 \\
 0.169810  & 0.0102 & 5.889 \\
 0.165874  & 0.0085 & 6.029 \\
\hline
 \end{tabular}
\end{table}

\subsection{Radial velocity modulation}\label{radvelmod}

The top panel of Fig.~\ref{lcrad} also shows the STELLA SES radial velocities from 2017. Note that a systematic zero point shift of 0.503\,\kms was added to this data, see \citet{2012AN....333..663S} for its determination with respect to the CORAVEL system. At this point we caution that the radial velocities in the \citet{2017A&A...600L...9J} paper differ in their zero point to same data re-plotted in the \citet{2018MNRAS.476.1140A}  paper by $\approx$5\,\kms .
A zero point correction of -2.6\,\kms should be applied for the \citet{2017A&A...600L...9J} data to satisfactorily fit our observations.

\begin{figure}[thb]
\centering
\includegraphics[angle=0,width=1.0\columnwidth]{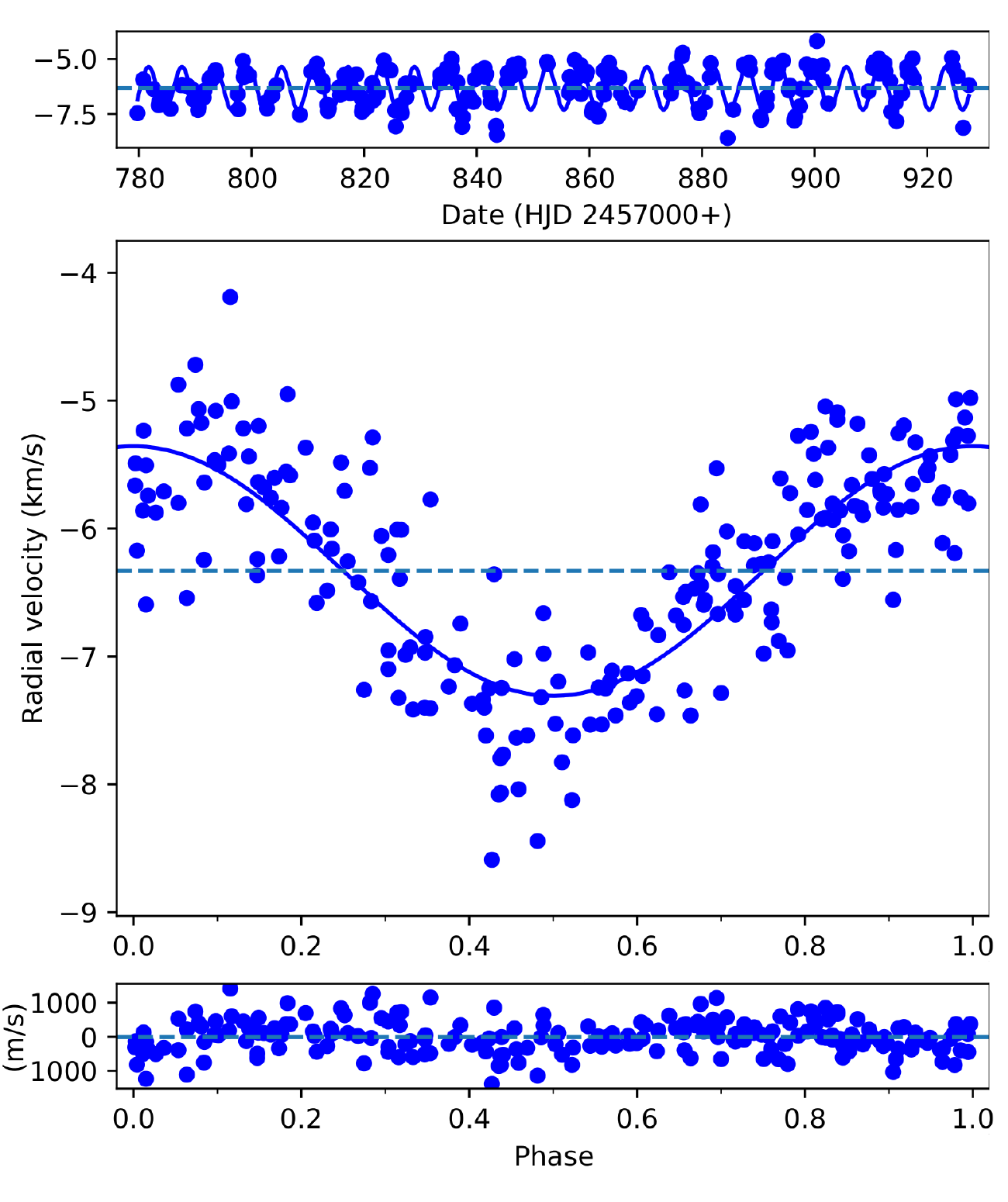}
\caption{Top: rotational modulation of the disk-integrated radial velocities of \ind. Middle: the radial velocity measurements are phase folded with the rotation period and fitted with a sinusoidal. Bottom: residuals of the sinusoidal fit.}
\label{SES_vrad}
\end{figure}

We first remove the grossly deviant velocities by a 3-$\sigma$ filter (this removed 9 data points and left 214). The remaining SES velocities of \inc have an internal precision of typically better than 2\,\kms. A Fourier analysis shows a clean peak at $f\approx$0.17, corresponding to a period of 5.95$\pm$0.03\,d, but with an almost as strong $1-f$ alias. Its full amplitude is almost 2\,\kms\ but with an rms of the sinusoidal fit of the same order (Fig.~\ref{SES_vrad}). Nevertheless, we can now confirm our earlier suggestion that the low-amplitude radial velocity jitter of \inc is spot modulated. Just recently, \citet{2018MNRAS.476.1140A} arrived at the same conclusion.

\section{Characteristics of the \Halpha profiles}\label{halp}

\subsection{Line-profile morphology}

\Halpha line profile variation is often associated with dynamo driven chromospheric activity, not seldom associated with a strong inhomogeneous stellar wind, coronal mass ejections and other violent events like flares. The most active stars have \Halpha in emission. Chromospheric, transition-region and coronal activity is indeed present in the case of \inc as amply demonstrated by, e.g. strong Ca\,{\sc ii} H\&K emission (see Paper~1), high-excitation UV lines like C\,{\sc iv} (Modigliani et al. 1993) as well as strong X-ray emission (Montez et al. 2010). The overall \Halpha emission line profile appears permanently asymmetric which suggests intense long-lived mass motions at the upper chromosphere, giving some support to an origin related to an active binary system with mass motions. Thus, one would expect some rotational modulation of it if it is related to \inc in the first place. 

\begin{figure}[]
\includegraphics[width=1.0\columnwidth]{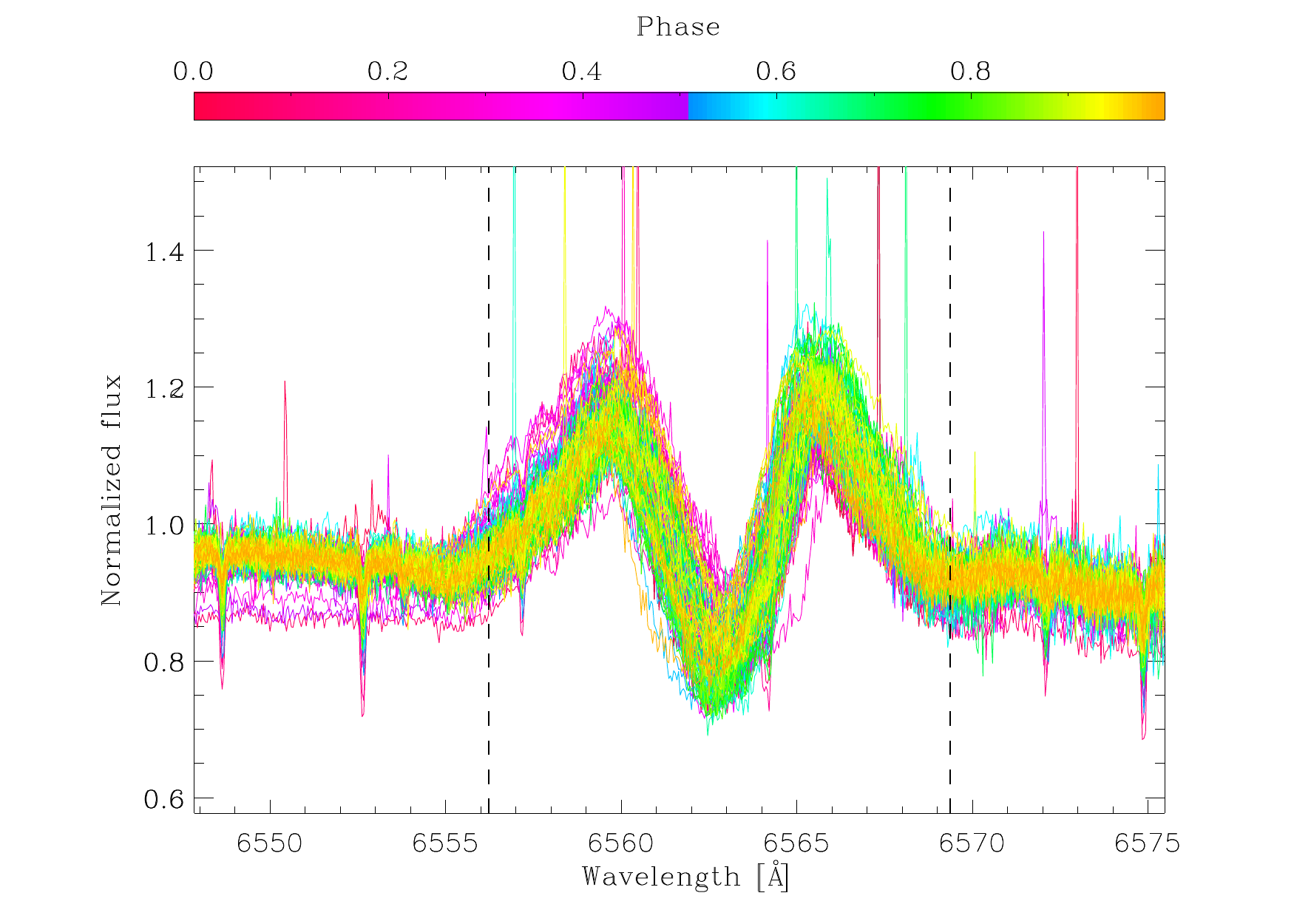}
\caption{Overplot of the \Halpha SES spectra of \inc from the first half of 2017. Indicated are the $\pm$300\,\kms\ period-search limits imposed for the 2D FFT in Fig.~\ref{Halpha2dfft}.}
\label{Halpha1d}
\end{figure}

A time series of 208 \Halpha line profiles from the first half of 2017 is shown in Fig.~\ref{Halpha1d}. A broad emission profile with an average FWHM of $\approx$400\,\kms\ (8.8~\AA ) superimposed with a central absorption reversal of width $\approx$150\,\kms\ appears consistently throughout the time series. The line width at continuum exceeds the expected rotational width by a factor of $\approx$5. The rotational period and the equatorial rotational velocity give a radius of 11\,R$_\odot$ (Table~\ref{astropars}, adopting an inclination of 45\degr\ from Doppler imaging). If the \Halpha emission is bound to the star above FWHM would then suggest an origin of at least part of the emission at an extended radius, e.g. due to a circumstellar environment of up to 3--4\,R$_\star$ (assuming corotation). This would mean that the overall \Halpha emission is composed of two parts; a chromospheric component and a circumstellar component. This has been seen and analyzed in several other (over)active stars, e.g. in UZ\,Librae \citep{2004A&A...421..295Z}, FK\,Comae \citep{1981ApJ...251L.101R}, II\,Pegasi \citep{1998A&A...338..191S} and others, and is not a specific issue for \inc because it is within a planetary nebula.

All three Balmer profiles of \inc vary slightly and consistently from one observation to the next while its profile morphology remains basically unaltered over our entire observing season. No rotational modulation was detected so far although there are changes seen in \Halpha on a decade-long scale \citep{2018MNRAS.476.1140A}. The two pseudo emission peaks frequently reverse its relative strength in our data set; once the blue emission is stronger once the red emission is stronger. At this point we note that the H$\beta$ and H$\gamma$ profiles of \inc look vastly different than \Halphad. Both are purely in absorption with an asymmetric shape but of same average width as the central \Halpha absorption. We measured FWHM for H$\beta$ of 150\,\kms and for H$\gamma$ of 160\,\kms; however, unfortunately, H$\beta$ falls at the edge of subsequent échelle-orders, while H$\gamma$ is significantly blended with a blue line at 4337.4\,\AA\ and therefore the S/N values for both lines are significanly lower compared with \Halphad; we estimate errors of 10-15\,\kms. No emission above the continuum is seen neither for H$\beta$ nor H$\gamma$. We note that the width of the \Halpha absorption reversal is in agreement with the expected rotational broadening, and so are the absorption profiles of H$\beta$ and H$\gamma$.

\subsection{Rotational modulation}

\begin{figure}[]
\includegraphics[width=0.95\columnwidth]{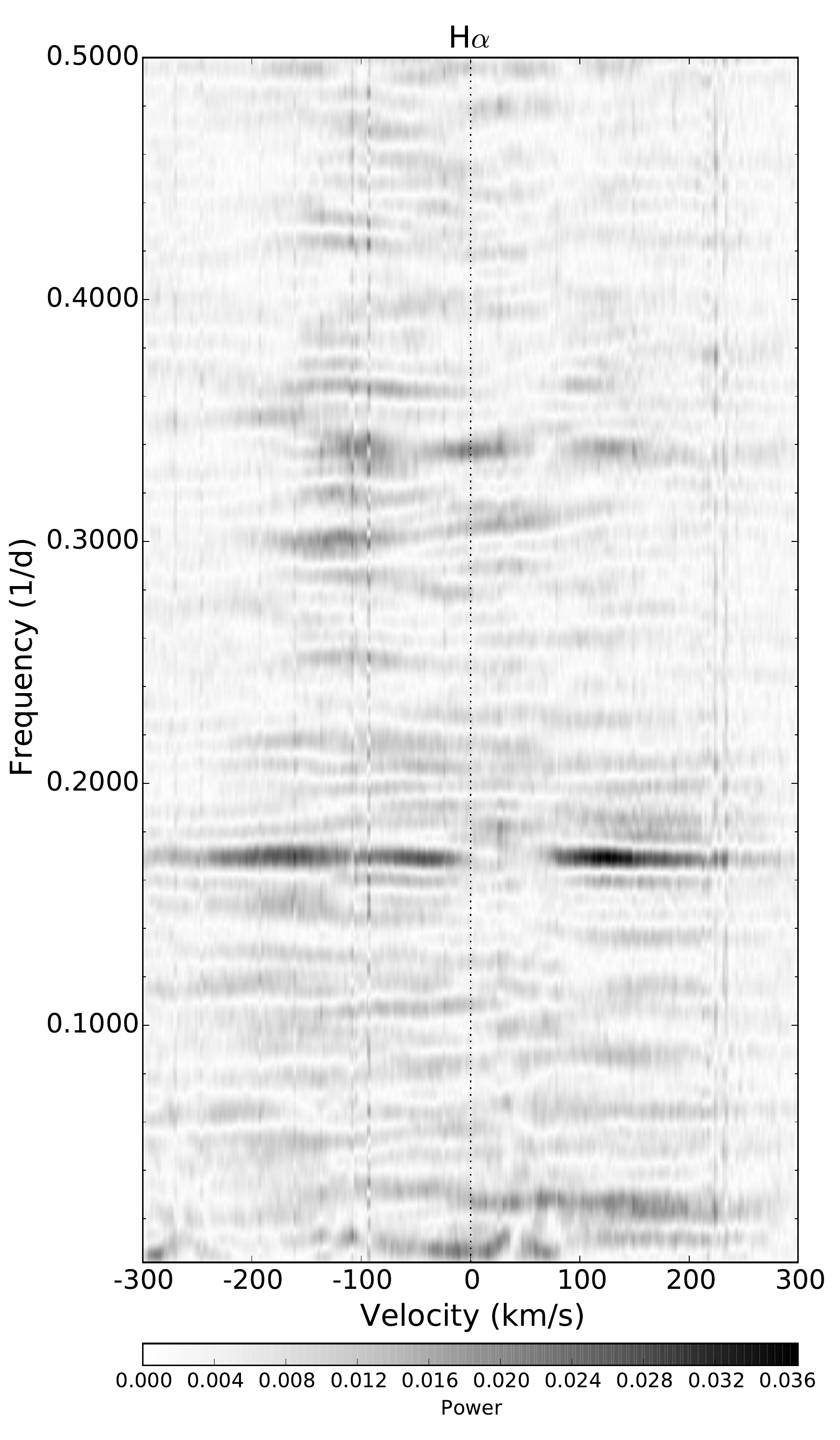}
\caption{Two-dimensional Fourier periodogram of the time series \Halpha profile from Fig.~\ref{Halpha1d}.  Spectral power is indicated in gray-scale. The plot's horizontal range is $\pm$300\,\kms\ around the line core while its vertical range is from 470\,d at the bottom to 2\,d at the top. The dominant excess power is detected at a frequency of $\approx$0.17, corresponding to a period of 5.92\,d. }
\label{Halpha2dfft}
\end{figure}

To search for coherent temporal changes, we apply a 2D Fourier-periodogram to the \Halpha time series. It is based on a simple fast Fourier transform (FFT) analysis to each wavelength-calibrated pixel of the \Halpha profile within a velocity range of $\pm$300~\kms\ around the line center \citep[for a more detailed description of the technique see][]{2014AN....335..904S}. Note that one SES CCD pixel disperses $\approx$0.06\,\AA\ at \Halphad. The resulting 2D periodogram is shown in Fig.~\ref{Halpha2dfft}. It reveals a clear and dominating peak at $f=0.169$~d$^{-1}$ ($P=5.92\pm0.06$\,d), i.e., the expected rotation period of the giant. A second, much weaker peak is detected at $2\,f$ and is identified as its alias.

It is puzzling though that our 2D periodogram shows a gap with zero power for the 0.169\,d$^{-1}$ frequency in the red part of the line core just between zero and $\approx$+70\,\kms\ velocity. We have no readily explanation for this.

\subsection{Line shape and width}

We compare the \inc \Halpha profiles with the chromospheric and transition-region models put forward by \citet{2004A&A...421..295Z}. With their model~4 (Table~3 and Fig.~4 in \citealt{2004A&A...421..295Z}), we find the overall best match for the average \inc profile. We note that the match is not based on a rigorous line-profile fit but only on a qualitative comparison. The enormous width of the \Halpha emission of $>400$\,\kms\ had been presented as a puzzle \citep{2018MNRAS.476.1140A} but is actually reproduced even with a normal plane-parallel atmosphere with an onset of the chromospheric temperature rise at around 5000\,K at a (solar-like) mass depth of $1$\,g\,cm$^{-2}$ and the assumption of complete frequency redistribution (CRD), which assumes that a photon absorbed in the wings is re-emitted in the core \citep[see][]{2003IAUS..210P.A21A}. The resulting \Halpha FWHM is of the order of 350\,\kms\ with a relative intensity of the emission peak of 1.2 and a 50\% central self reversal. It implies an upper chromosphere with a temperature of 10,000\,K and a logarithmic column density of $-2.7$ as well as a transition region (to the corona) with a temperature in excess of 100,000\,K and a logarithmic column density of $-6$.

\section{Summary and discussion}\label{disc}

\begin{figure}[]
\includegraphics[width=1\columnwidth]{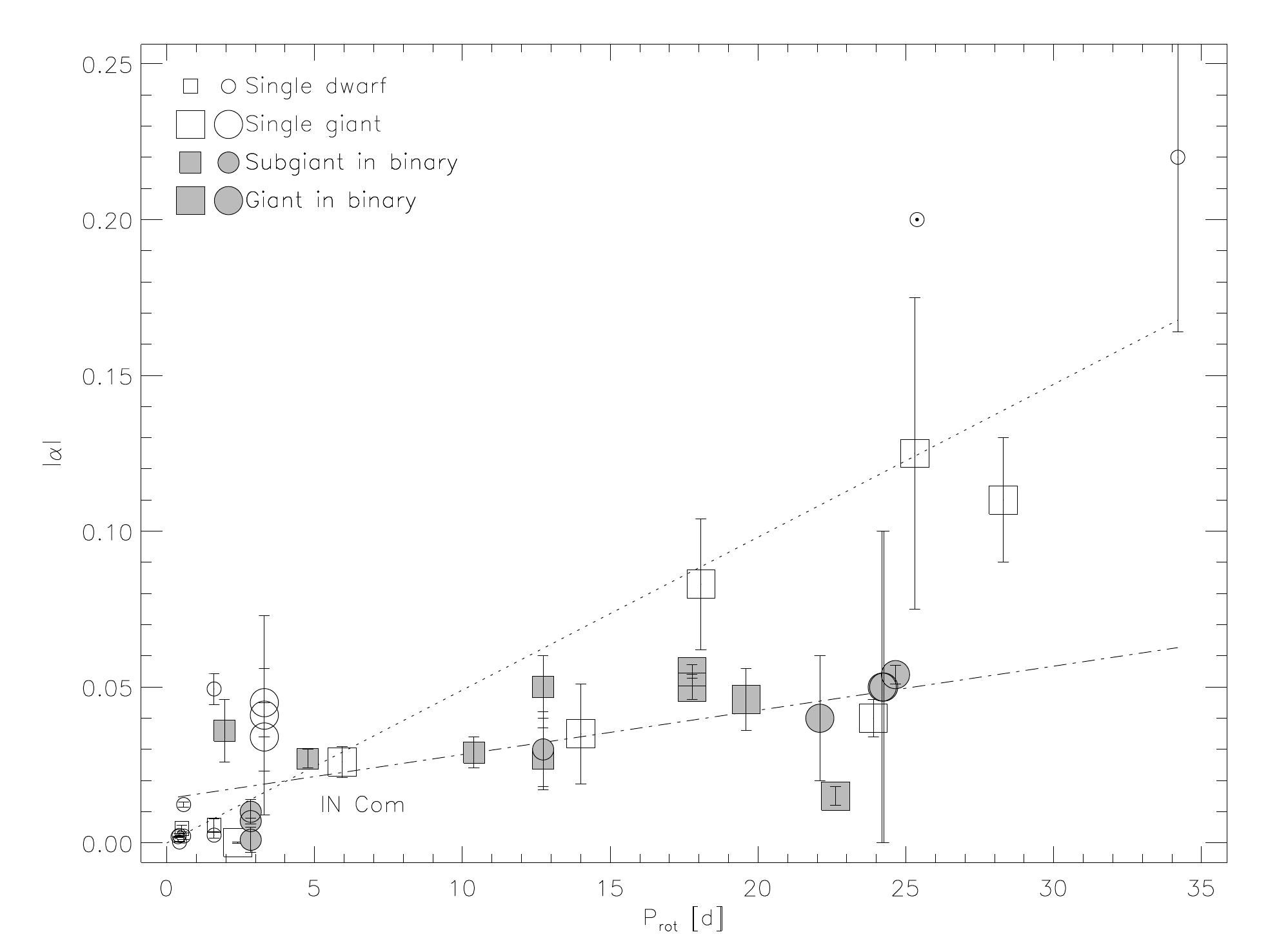}
\caption{Relationships between rotation and differential rotation for late-type single and binary stars. The position for \inc is in agreement with the linear fit to (effectively) single stars, represented by the dotted line, suggesting that $|\alpha|$\,$\propto$\,$P_{\rm rot}$[d]/200.}\label{rdrplot}
\end{figure}

We have analyzed decade-long photometric and one season-long spectroscopic data of \inc to derive more accurate stellar parameters and perform a time-series Doppler imaging study.
From the long-term photometric observations we have confirmed a $\approx$5.973 day-long equatorial rotation period of the G-star. Also, we have provided more accurate astrophysical parameters for \ind. Our time-series Doppler imaging study for the first half of 2017 yielded 13 subsequent surface image reconstructions, which were used to estimate surface differential rotation. We found antisolar surface rotation profile with $\alpha=-0.026$ shear coefficient.
This value falls within the recently proposed rotation-differential rotation relationship by \citet{2017AN....338..903K}, see Fig.~\ref{rdrplot}. According to the plot, the linear fit for (effectively) single stars suggests $|\alpha|$\,$\propto$\,$P_{\rm rot}$[d]/200. Moreover, the derived absolute surface shear of $\Delta\Omega=0.027$[rad/d] would follow the general trend of $\Delta\Omega\propto T_{\rm eff}^{p}$ where $p=5.8\pm1.0$ \citep[see Fig.~2 in][]{2017AN....338..903K}. Indeed, from the long-term photometric period variations we estimated the rate of the surface shear to be $\Delta P/P\approx0.03$, which was in agreement with the shear coefficient derived from Doppler imaging.

The G-giant star in the center of the planetary nebula shows features originating from its evolutionary history. The giant is a barium-rich star; from our spectral synthesis (see Sect.~\ref{astrop}) we estimate a [Ba/Fe] ratio of 0.85$\pm$0.25, supporting the former result of 0.50$\pm$0.30 by \citet{1997A&A...320..913T}. We redetermined [Y/Fe] and [Sr/Fe] ratios as well, confirming the overabundance of these elements in the atmosphere of \ind. The present configuration and the overabundant s-process elements of the G-star in the binary could be explained if the precursor of the white dwarf had originally the higher mass and therefore evolved faster to the white dwarf stage, while losing mass, and afterwards the companion (now G-star) was polluted by mass-transfer or wind accretion \citep[cf.][]{1995A&A...296..709V}.

The parallel variation of the long-term light curve with the orbital phase suggests a connection between the orbital motion of the binary and the activity of the G-giant. The star is the brightest and faintest at minimum and maximum radial velocity, respectively (see Sect.~\ref{phot} Fig.~\ref{lcrad}).

We note finally that, although no observational evidence has been found so far to support the existence of accreting material around the G-star, its high angular momentum, the peculiar differential rotation, the H$\alpha$ behaviour and the parallel variation of the long-term brightness with the orbital phase, may all be explained by the presence of an accretion disc tilted to the orbit.

\begin{acknowledgements}
We thank our anonymous referee for their valuable suggestions that have helped to improve the paper. This paper is based on data obtained with the STELLA robotic telescopes in Tenerife, an AIP facility jointly operated by AIP and IAC (https://stella.aip.de/) and by the Amadeus APT jointly operated by AIP and Fairborn Observatory in Arizona. For their continuous support, we are grateful to the ministry for research and culture of the State of Brandenburg (MWFK) and the German federal ministry for education and research (BMBF).
Authors from Konkoly Observatory acknowledge support from the Austrian-Hungarian Action Foundation (OMAA). KV is grateful to the Hungarian National Research, Development and Innovation Office for OTKA grant K-113117. KV is supported by the Bolyai J\'anos Research Scholarship of the Hungarian Academy of Sciences.
The authors acknowledge the support of the German \emph{Deut\-sche For\-schungs\-ge\-mein\-schaft, DFG\/} through projects KO2320/1 and STR645/1.
This work has made use of data from the European Space Agency (ESA)
mission {\it Gaia} (\url{https://www.cosmos.esa.int/gaia}), processed by
the {\it Gaia} Data Processing and Analysis Consortium (DPAC,
\url{https://www.cosmos.esa.int/web/gaia/dpac/consortium}). Funding
for the DPAC has been provided by national institutions, in particular
the institutions participating in the {\it Gaia} Multilateral Agreement.
\end{acknowledgements}

\bibliography{kovarietal_incom}

\begin{thebibliography}{57}
\expandafter\ifx\csname natexlab\endcsname\relax\def\natexlab#1{#1}\fi

\bibitem[{{Aller} {et~al.}(2018){Aller}, {Lillo-Box}, {Vu{\v c}kovi{\'c}}, {Van
  Winckel}, {Jones}, {Montesinos}, {Zorotovic}, \&
  {Miranda}}]{2018MNRAS.476.1140A}
{Aller}, A., {Lillo-Box}, J., {Vu{\v c}kovi{\'c}}, M., {et~al.} 2018, \mnras,
  476, 1140

\bibitem[{{Avrett} \& {Loeser}(2003)}]{2003IAUS..210P.A21A}
{Avrett}, E.~H. \& {Loeser}, R. 2003, in IAU Symposium, Vol. 210, Modelling of
  Stellar Atmospheres, ed. N.~{Piskunov}, W.~W. {Weiss}, \& D.~F. {Gray}, A21

\bibitem[{{Bisterzo} {et~al.}(2011){Bisterzo}, {Gallino}, {Straniero},
  {Cristallo}, \& {K{\"a}ppeler}}]{2011MNRAS.418..284B}
{Bisterzo}, S., {Gallino}, R., {Straniero}, O., {Cristallo}, S., \&
  {K{\"a}ppeler}, F. 2011, \mnras, 418, 284

\bibitem[{{Carroll} {et~al.}(2008){Carroll}, {Kopf}, \&
  {Strassmeier}}]{2008A&A...488..781C}
{Carroll}, T.~A., {Kopf}, M., \& {Strassmeier}, K.~G. 2008, \aap, 488, 781

\bibitem[{{Carroll} {et~al.}(2012){Carroll}, {Strassmeier}, {Rice}, \&
  {K{\"u}nstler}}]{2012A&A...548A..95C}
{Carroll}, T.~A., {Strassmeier}, K.~G., {Rice}, J.~B., \& {K{\"u}nstler}, A.
  2012, \aap, 548, A95

\bibitem[{{Castelli} \& {Kurucz}(2004)}]{2004astro.ph..5087C}
{Castelli}, F. \& {Kurucz}, R.~L. 2004, ArXiv Astrophysics e-prints

\bibitem[{{Ciardullo} {et~al.}(1999){Ciardullo}, {Bond}, {Sipior}, {Fullton},
  {Zhang}, \& {Schaefer}}]{1999AJ....118..488C}
{Ciardullo}, R., {Bond}, H.~E., {Sipior}, M.~S., {et~al.} 1999, \aj, 118, 488

\bibitem[{{Csubry} \& {Koll{\'a}th}(2004)}]{2004ESASP.559..396C}
{Csubry}, Z. \& {Koll{\'a}th}, Z. 2004, in ESA Special Publication, Vol. 559,
  SOHO 14 Helio- and Asteroseismology: Towards a Golden Future, ed.
  D.~{Danesy}, 396

\bibitem[{{De Marco} {et~al.}(2004){De Marco}, {Bond}, {Harmer}, \&
  {Fleming}}]{2004ApJ...602L..93D}
{De Marco}, O., {Bond}, H.~E., {Harmer}, D., \& {Fleming}, A.~J. 2004, \apjl,
  602, L93

\bibitem[{{Donati} \& {Collier Cameron}(1997)}]{1997MNRAS.291....1D}
{Donati}, J.-F. \& {Collier Cameron}, A. 1997, \mnras, 291, 1

\bibitem[{{Feibelman} \& {Kaler}(1983)}]{1983ApJ...269..592F}
{Feibelman}, W.~A. \& {Kaler}, J.~B. 1983, \apj, 269, 592

\bibitem[{{Flower}(1996)}]{1996ApJ...469..355F}
{Flower}, P.~J. 1996, \apj, 469, 355

\bibitem[{{Gaia Collaboration} {et~al.}(2018){Gaia Collaboration}, {Brown},
  {Vallenari}, {Prusti}, {de Bruijne}, {Babusiaux}, \&
  {Bailer-Jones}}]{2018arXiv180409365G}
{Gaia Collaboration}, {Brown}, A.~G.~A., {Vallenari}, A., {et~al.} 2018, ArXiv
  e-prints

\bibitem[{{Gonz{\'a}lez Mart{\'{\i}}nez-Pa{\'{\i}}s}
  {et~al.}(2014){Gonz{\'a}lez Mart{\'{\i}}nez-Pa{\'{\i}}s}, {Shahbaz}, \&
  {Casares Vel{\'a}zquez}}]{2014apa..book.....G}
{Gonz{\'a}lez Mart{\'{\i}}nez-Pa{\'{\i}}s}, I., {Shahbaz}, T., \& {Casares
  Vel{\'a}zquez}, J. 2014, {Accretion Processes in Astrophysics}

\bibitem[{{Granzer} {et~al.}(2001){Granzer}, {Reegen}, \&
  {Strassmeier}}]{2001AN....322..325G}
{Granzer}, T., {Reegen}, P., \& {Strassmeier}, K.~G. 2001, Astronomische
  Nachrichten, 322, 325

\bibitem[{{Gray}(1981)}]{1981ApJ...251..155G}
{Gray}, D.~F. 1981, \apj, 251, 155

\bibitem[{{Gray} \& {Toner}(1986)}]{1986ApJ...310..277G}
{Gray}, D.~F. \& {Toner}, C.~G. 1986, \apj, 310, 277

\bibitem[{{Guerrero}(2012)}]{2012IAUS..283..204G}
{Guerrero}, M.~A. 2012, in IAU Symposium, Vol. 283, IAU Symposium, 204--210

\bibitem[{{Gustafsson} {et~al.}(2008){Gustafsson}, {Edvardsson}, {Eriksson},
  {J{\o}rgensen}, {Nordlund}, \& {Plez}}]{2008A&A...486..951G}
{Gustafsson}, B., {Edvardsson}, B., {Eriksson}, K., {et~al.} 2008, \aap, 486,
  951

\bibitem[{{Jasniewicz} {et~al.}(1987){Jasniewicz}, {Acker}, \&
  {Duquennoy}}]{1987A&A...180..145J}
{Jasniewicz}, G., {Acker}, A., \& {Duquennoy}, A. 1987, \aap, 180, 145

\bibitem[{{Jasniewicz} {et~al.}(1994){Jasniewicz}, {Acker}, {Mauron},
  {Duquennoy}, \& {Cuypers}}]{1994A&A...286..211J}
{Jasniewicz}, G., {Acker}, A., {Mauron}, N., {Duquennoy}, A., \& {Cuypers}, J.
  1994, \aap, 286, 211

\bibitem[{{Jasniewicz} {et~al.}(1996){Jasniewicz}, {Thevenin}, {Monier}, \&
  {Skiff}}]{1996A&A...307..200J}
{Jasniewicz}, G., {Thevenin}, F., {Monier}, R., \& {Skiff}, B.~A. 1996, \aap,
  307, 200

\bibitem[{{Jones} \& {Boffin}(2017)}]{2017NatAs...1E.117J}
{Jones}, D. \& {Boffin}, H.~M.~J. 2017, Nature Astronomy, 1, 0117

\bibitem[{{Jones} {et~al.}(2017){Jones}, {Van Winckel}, {Aller}, {Exter}, \&
  {De Marco}}]{2017A&A...600L...9J}
{Jones}, D., {Van Winckel}, H., {Aller}, A., {Exter}, K., \& {De Marco}, O.
  2017, \aap, 600, L9

\bibitem[{{K{\H o}v{\'a}ri} {et~al.}(2015){K{\H o}v{\'a}ri}, {Kriskovics},
  {K{\"u}nstler}, {Carroll}, {Strassmeier}, {Vida}, {Ol{\'a}h}, {Bartus}, \&
  {Weber}}]{2015A&A...573A..98K}
{K{\H o}v{\'a}ri}, {\mbox{Zs}}., {Kriskovics}, L., {K{\"u}nstler}, A., {et~al.}
  2015, \aap, 573, A98

\bibitem[{{K{\H o}v{\'a}ri} {et~al.}(2017){K{\H o}v{\'a}ri}, {Ol{\'a}h},
  {Kriskovics}, {Vida}, {Forg{\'a}cs-Dajka}, \&
  {Strassmeier}}]{2017AN....338..903K}
{K{\H o}v{\'a}ri}, {\mbox{Zs}}., {Ol{\'a}h}, K., {Kriskovics}, L., {et~al.}
  2017, Astronomische Nachrichten, 338, 903

\bibitem[{{Kuczawska} \& {Mikolajewski}(1993)}]{1993AcA....43..445K}
{Kuczawska}, E. \& {Mikolajewski}, M. 1993, \actaa, 43, 445

\bibitem[{{K{\"u}nstler} {et~al.}(2015){K{\"u}nstler}, {Carroll}, \&
  {Strassmeier}}]{2015A&A...578A.101K}
{K{\"u}nstler}, A., {Carroll}, T.~A., \& {Strassmeier}, K.~G. 2015, \aap, 578,
  A101

\bibitem[{{Kupka} {et~al.}(1999){Kupka}, {Piskunov}, {Ryabchikova}, {Stempels},
  \& {Weiss}}]{1999A&AS..138..119K}
{Kupka}, F., {Piskunov}, N., {Ryabchikova}, T.~A., {Stempels}, H.~C., \&
  {Weiss}, W.~W. 1999, \aaps, 138, 119

\bibitem[{{Lindborg} {et~al.}(2014){Lindborg}, {Hackman}, {Mantere},
  {Korhonen}, {Ilyin}, {Kochukhov}, \& {Piskunov}}]{2014A&A...562A.139L}
{Lindborg}, M., {Hackman}, T., {Mantere}, M.~J., {et~al.} 2014, \aap, 562, A139

\bibitem[{{Longmore} \& {Tritton}(1980)}]{1980MNRAS.193..521L}
{Longmore}, A.~J. \& {Tritton}, S.~B. 1980, \mnras, 193, 521

\bibitem[{{Malasan} {et~al.}(1991){Malasan}, {Yamasaki}, \&
  {Kondo}}]{1991AJ....101.2131M}
{Malasan}, H.~L., {Yamasaki}, A., \& {Kondo}, M. 1991, \aj, 101, 2131

\bibitem[{{Montez} {et~al.}(2010){Montez}, {De Marco}, {Kastner}, \&
  {Chu}}]{2010ApJ...721.1820M}
{Montez}, Jr., R., {De Marco}, O., {Kastner}, J.~H., \& {Chu}, Y.-H. 2010,
  \apj, 721, 1820

\bibitem[{{Noskova}(1989)}]{1989SvAL...15..149N}
{Noskova}, R.~I. 1989, Soviet Astronomy Letters, 15, 149

\bibitem[{{Paczy\'nski}(1976)}]{1976IAUS...73...75P}
{Paczy\'nski}, B. 1976, in IAU Symposium, Vol.~73, Structure and Evolution of
  Close Binary Systems, ed. P.~{Eggleton}, S.~{Mitton}, \& J.~{Whelan}, 75

\bibitem[{{Piskunov} \& {Valenti}(2017)}]{2017A&A...597A..16P}
{Piskunov}, N. \& {Valenti}, J.~A. 2017, \aap, 597, A16

\bibitem[{{Podsiadlowski}(2001)}]{2001ASPC..229..239P}
{Podsiadlowski}, P. 2001, in Astronomical Society of the Pacific Conference
  Series, Vol. 229, Evolution of Binary and Multiple Star Systems, ed.
  P.~{Podsiadlowski}, S.~{Rappaport}, A.~R. {King}, F.~{D'Antona}, \&
  L.~{Burderi}, 239

\bibitem[{{Pojmanski}(2002)}]{2002AcA....52..397P}
{Pojmanski}, G. 2002, \actaa, 52, 397

\bibitem[{{Ramsey} {et~al.}(1981){Ramsey}, {Nations}, \&
  {Barden}}]{1981ApJ...251L.101R}
{Ramsey}, L.~W., {Nations}, H.~L., \& {Barden}, S.~C. 1981, \apjl, 251, L101

\bibitem[{{Short} {et~al.}(1998){Short}, {Byrne}, \&
  {Panagi}}]{1998A&A...338..191S}
{Short}, C.~I., {Byrne}, P.~B., \& {Panagi}, P.~M. 1998, \aap, 338, 191

\bibitem[{{Strassmeier} {et~al.}(1997{\natexlab{a}}){Strassmeier}, {Boyd},
  {Epand}, \& {Granzer}}]{1997PASP..109..697S}
{Strassmeier}, K.~G., {Boyd}, L.~J., {Epand}, D.~H., \& {Granzer}, T.
  1997{\natexlab{a}}, \pasp, 109, 697

\bibitem[{{Strassmeier} {et~al.}(2010){Strassmeier}, {Granzer}, {Weber},
  {Woche}, {Popow}, {J{\"a}rvinen}, {Bartus}, {Bauer}, {Dionies}, {Fechner},
  {Bittner}, \& {Paschke}}]{2010AdAst2010E..19S}
{Strassmeier}, K.~G., {Granzer}, T., {Weber}, M., {et~al.} 2010, Advances in
  Astronomy, 2010, 19

\bibitem[{{Strassmeier} {et~al.}(1997{\natexlab{b}}){Strassmeier}, {Hubl}, \&
  {Rice}}]{1997A&A...322..511S}
{Strassmeier}, K.~G., {Hubl}, B., \& {Rice}, J.~B. 1997{\natexlab{b}}, \aap,
  322, 511

\bibitem[{{Strassmeier} {et~al.}(2015){Strassmeier}, {Ilyin}, {J{\"a}rvinen},
  {Weber}, {Woche}, {Barnes}, {Bauer}, {Beckert}, {Bittner}, {Bredthauer},
  {Carroll}, {Denker}, {Dionies}, {DiVarano}, {D{\"o}scher}, {Fechner},
  {Feuerstein}, {Granzer}, {Hahn}, {Harnisch}, {Hofmann}, {Lesser}, {Paschke},
  {Pankratow}, {Plank}, {Pl{\"u}schke}, {Popow}, \&
  {Sablowski}}]{2015AN....336..324S}
{Strassmeier}, K.~G., {Ilyin}, I., {J{\"a}rvinen}, A., {et~al.} 2015,
  Astronomische Nachrichten, 336, 324

\bibitem[{{Strassmeier} {et~al.}(2018){Strassmeier}, {Ilyin}, \&
  {Steffen}}]{2018A&A...612A..44S}
{Strassmeier}, K.~G., {Ilyin}, I., \& {Steffen}, M. 2018, \aap, 612, A44

\bibitem[{{Strassmeier} \& {Ol{\'a}h}(2004)}]{2004ESASP.538..149S}
{Strassmeier}, K.~G. \& {Ol{\'a}h}, K. 2004, in ESA Special Publication, Vol.
  538, Stellar Structure and Habitable Planet Finding, ed. F.~{Favata},
  S.~{Aigrain}, \& A.~{Wilson}, 149--161

\bibitem[{{Strassmeier} {et~al.}(2012){Strassmeier}, {Weber}, {Granzer}, \&
  {J{\"a}rvinen}}]{2012AN....333..663S}
{Strassmeier}, K.~G., {Weber}, M., {Granzer}, T., \& {J{\"a}rvinen}, S. 2012,
  Astronomische Nachrichten, 333, 663

\bibitem[{{Strassmeier} {et~al.}(2014){Strassmeier}, {Weber}, {Granzer},
  {Schanne}, {Bartus}, \& {Ilyin}}]{2014AN....335..904S}
{Strassmeier}, K.~G., {Weber}, M., {Granzer}, T., {et~al.} 2014, Astronomische
  Nachrichten, 335, 904

\bibitem[{{Th\'evenin} \& {Jasniewicz}(1997)}]{1997A&A...320..913T}
{Th\'evenin}, F. \& {Jasniewicz}, G. 1997, \aap, 320, 913

\bibitem[{{Tout} \& {Reg{\H o}s}(2003)}]{2003ASPC..293..100T}
{Tout}, C.~A. \& {Reg{\H o}s}, E. 2003, in Astronomical Society of the Pacific
  Conference Series, Vol. 293, 3D Stellar Evolution, ed. S.~{Turcotte}, S.~C.
  {Keller}, \& R.~M. {Cavallo}, 100

\bibitem[{{Van Winckel} {et~al.}(2014){Van Winckel}, {Jorissen}, {Exter},
  {Raskin}, {Prins}, {Perez Padilla}, {Merges}, \&
  {Pessemier}}]{2014A&A...563L..10V}
{Van Winckel}, H., {Jorissen}, A., {Exter}, K., {et~al.} 2014, \aap, 563, L10

\bibitem[{{Verbunt} \& {Phinney}(1995)}]{1995A&A...296..709V}
{Verbunt}, F. \& {Phinney}, E.~S. 1995, \aap, 296, 709

\bibitem[{{Vida} {et~al.}(2014){Vida}, {Ol{\'a}h}, \&
  {Szab{\'o}}}]{2014MNRAS.441.2744V}
{Vida}, K., {Ol{\'a}h}, K., \& {Szab{\'o}}, R. 2014, \mnras, 441, 2744

\bibitem[{{Weber} {et~al.}(2012){Weber}, {Granzer}, \&
  {Strassmeier}}]{2012SPIE.8451E..0KW}
{Weber}, M., {Granzer}, T., \& {Strassmeier}, K.~G. 2012, in Society of
  Photo-Optical Instrumentation Engineers (SPIE) Conference Series, Vol. 8451,
  Society of Photo-Optical Instrumentation Engineers (SPIE) Conference Series,
  0

\bibitem[{{Weber} {et~al.}(2008){Weber}, {Granzer}, {Strassmeier}, \&
  {Woche}}]{2008SPIE.7019E..0LW}
{Weber}, M., {Granzer}, T., {Strassmeier}, K.~G., \& {Woche}, M. 2008, in
  Society of Photo-Optical Instrumentation Engineers (SPIE) Conference Series,
  Vol. 7019, Society of Photo-Optical Instrumentation Engineers (SPIE)
  Conference Series, 0

\bibitem[{{Weber} \& {Strassmeier}(2011)}]{2011A&A...531A..89W}
{Weber}, M. \& {Strassmeier}, K.~G. 2011, \aap, 531, A89

\bibitem[{{Zboril} {et~al.}(2004){Zboril}, {Strassmeier}, \&
  {Avrett}}]{2004A&A...421..295Z}
{Zboril}, M., {Strassmeier}, K.~G., \& {Avrett}, E.~H. 2004, \aap, 421, 295

\end{thebibliography}
\bibliographystyle{aa}

%
%
\begin{appendix}
\twocolumn
\section{Log of spectroscopic data}

\vspace{0.6cm}
\begin{center}
\tablehead{\hline\hline\noalign{\smallskip}
\setcounter{footnote}{0}HJD\footnote{a}&Phase\footnote{b}\setcounter{footnote}{-1}&  Date  & S/N & Subset\\
\hline\noalign{\smallskip}}
\topcaption{Observing log of STELLA-SES spectra of \inc from 2017 used for individual Doppler reconstructions shown in Sect.~\ref{di}}\label{Tab1}
\begin{supertabular}{c c c r c}
\tabletail{\hline\hline\noalign{\smallskip}
\multicolumn{5}{l}{\setcounter{footnote}{0}\footnote{a}{2\,450\,000+}}\\
\multicolumn{5}{l}{\footnote{b}{Phases computed using Eq.~1.}}\\}
\tablelasttail{
\hline\hline\noalign{\smallskip}
\multicolumn{5}{l}{\setcounter{footnote}{0}\footnote{a}{2\,450\,000+}}\\
\multicolumn{5}{l}{\footnote{b}{Phases computed using Eq.~1.}}}
7780.746 &	0.799 &	27.01.2017 &	102 & S01 \\
7782.600 &	0.111 &	29.01.2017 &	184 & S01 \\
7783.521 &	0.266 &	29.01.2017 &	149 & S01 \\
7783.676 &	0.292 &	30.01.2017 &	170 & S01 \\
7784.611 &	0.450 &	31.01.2017 &	160 & S01 \\
7785.603 &	0.617 &	01.02.2017 &	140 & S01 \\
7787.503 &	0.937 &	02.02.2017 &	144 & S01 \\
7787.656 &	0.963 &	03.02.2017 &	176 & S01 \\
7788.653 &	0.131 &	04.02.2017 &	166 & S02 \\
7789.502 &	0.274 &	04.02.2017 &	135 & S02 \\
7790.496 &	0.442 &	05.02.2017 &	144 & S02 \\
7790.617 &	0.462 &	06.02.2017 &	195 & S02 \\
7791.498 &	0.611 &	06.02.2017 &	143 & S02 \\
7791.618 &	0.631 &	07.02.2017 &	185 & S02 \\
7792.498 &	0.779 &	07.02.2017 &	147 & S02 \\
7792.619 &	0.800 &	08.02.2017 &	185 & S02 \\
7793.499 &	0.948 &	08.02.2017 &	169 & S02 \\
7793.620 &	0.968 &	09.02.2017 &	197 & S02 \\
7793.744 &	0.989 &	09.02.2017 &	188 & S02 \\
7808.623 &	0.497 &	24.02.2017 &	186 & S03 \\
7811.607 &	0.999 &	27.02.2017 &	178 & S03 \\
7811.730 &	0.020 &	27.02.2017 &	168 & S03 \\
7812.494 &	0.149 &	27.02.2017 &	175 & S03 \\
7812.618 &	0.170 &	28.02.2017 &	193 & S03 \\
7812.738 &	0.190 &	28.02.2017 &	174 & S03 \\
7813.490 &	0.317 &	28.02.2017 &	177 & S03 \\
7813.612 &	0.337 &	01.03.2017 &	198 & S03 \\
7813.733 &	0.358 &	01.03.2017 &	186 & S03 \\
7815.593 &	0.671 &	03.03.2017 &	191 & S03 \\
7815.715 &	0.692 &	03.03.2017 &	137 & S03 \\
7816.486 &	0.822 &	03.03.2017 &	154 & S03 \\
7816.607 &	0.842 &	04.03.2017 &	184 & S03 \\
7817.506 &	0.994 &	04.03.2017 &	105 & S04 \\
7817.628 &	0.014 &	05.03.2017 &	110 & S04 \\
7818.495 &	0.160 &	05.03.2017 &	153 & S04 \\
7818.616 &	0.181 &	06.03.2017 &	123 & S04 \\
7819.431 &	0.318 &	06.03.2017 &	118 & S04 \\
7819.599 &	0.346 &	07.03.2017 &	181 & S04 \\
7819.720 &	0.367 &	07.03.2017 &	134 & S04 \\
7820.481 &	0.495 &	07.03.2017 &	108 & S04 \\
7821.547 &	0.675 &	09.03.2017 &	116 & S04 \\
7821.668 &	0.695 &	09.03.2017 &	142 & S04 \\
7822.473 &	0.831 &	09.03.2017 &	101 & S04 \\
7823.490 &	0.002 &	10.03.2017 &	177 & S05 \\
7823.603 &	0.021 &	11.03.2017 &	132 & S05 \\
7824.491 &	0.171 &	11.03.2017 &	166 & S05 \\
7824.694 &	0.205 &	12.03.2017 &	 90 & S05 \\
7825.490 &	0.339 &	12.03.2017 &	169 & S05 \\
7825.615 &	0.360 &	13.03.2017 &	119 & S05 \\
7826.397 &	0.492 &	13.03.2017 &	117 & S05 \\
7826.521 &	0.513 &	13.03.2017 &	156 & S05 \\
7826.711 &	0.545 &	14.03.2017 &	173 & S05 \\
7827.397 &	0.660 &	14.03.2017 &	122 & S05 \\
7827.520 &	0.681 &	14.03.2017 &	184 & S05 \\
7828.391 &	0.828 &	15.03.2017 &	 97 & S05 \\
7828.720 &	0.883 &	16.03.2017 &	165 & S05 \\
7833.669 &	0.717 &	21.03.2017 &	164 & S06 \\
7834.401 &	0.841 &	21.03.2017 &	144 & S06 \\
7834.521 &	0.861 &	21.03.2017 &	178 & S06 \\
7834.676 &	0.887 &	22.03.2017 &	147 & S06 \\
7835.524 &	0.030 &	22.03.2017 &	143 & S06 \\
7835.644 &	0.050 &	23.03.2017 &	146 & S06 \\
7836.696 &	0.228 &	24.03.2017 &	145 & S06 \\
7837.524 &	0.367 &	24.03.2017 &	175 & S06 \\
7838.401 &	0.515 &	25.03.2017 &	167 & S06 \\
7838.523 &	0.535 &	25.03.2017 &	183 & S07 \\
7839.432 &	0.689 &	26.03.2017 &	 88 & S07 \\
7840.402 &	0.852 &	27.03.2017 &	168 & S07 \\
7840.525 &	0.873 &	27.03.2017 &	169 & S07 \\
7840.647 &	0.893 &	28.03.2017 &	171 & S07 \\
7841.404 &	0.021 &	28.03.2017 &	155 & S07 \\
7841.527 &	0.042 &	28.03.2017 &	178 & S07 \\
7841.651 &	0.062 &	29.03.2017 &	169 & S07 \\
7842.404 &	0.189 &	29.03.2017 &	159 & S07 \\
7842.525 &	0.210 &	29.03.2017 &	169 & S07 \\
7842.646 &	0.230 &	30.03.2017 &	130 & S07 \\
7843.440 &	0.364 &	30.03.2017 &	148 & S07 \\
7843.573 &	0.386 &	31.03.2017 &	 91 & S07 \\
7860.400 &	0.222 &	16.04.2017 &	164 & S08 \\
7861.400 &	0.391 &	17.04.2017 &	173 & S08 \\
7861.645 &	0.432 &	18.04.2017 &	161 & S08 \\
7863.401 &	0.728 &	19.04.2017 &	164 & S08 \\
7864.401 &	0.896 &	20.04.2017 &	 87 & S08 \\
7864.650 &	0.938 &	21.04.2017 &	149 & S08 \\
7865.623 &	0.102 &	22.04.2017 &	146 & S08 \\
7868.403 &	0.571 &	24.04.2017 &	186 & S08\\
7868.566 &	0.598 &	25.04.2017 &	163 & S08 \\
7874.406 &	0.582 &	30.04.2017 &	180 & S09 \\
7874.526 &	0.603 &	30.04.2017 &	159 & S09 \\
7874.699 &	0.632 &	01.05.2017 &	120 & S09 \\
7875.406 &	0.751 &	01.05.2017 &	171 & S09 \\
7876.450 &	0.927 &	02.05.2017 &	159 & S09 \\
7877.531 &	0.109 &	03.05.2017 &	141 & S09 \\
7879.404 &	0.425 &	05.05.2017 &	170 & S09 \\
7879.525 &	0.445 &	05.05.2017 &	167 & S09 \\
7884.555 &	0.293 &	11.05.2017 &	 92 & S09 \\
7890.413 &	0.280 &	16.05.2017 &	112 & S10 \\
7890.534 &	0.300 &	17.05.2017 &	128 & S10 \\
7891.414 &	0.449 &	17.05.2017 &	150 & S10 \\
7891.534 &	0.469 &	18.05.2017 &	150 & S10 \\
7892.484 &	0.629 &	18.05.2017 &	148 & S10 \\
7892.606 &	0.650 &	19.05.2017 &	120 & S10 \\
7893.414 &	0.786 &	19.05.2017 &	142 & S10 \\
7893.535 &	0.806 &	20.05.2017 &	162 & S10 \\
7894.415 &	0.954 &	20.05.2017 &	107 & S10 \\
7895.415 &	0.123 &	21.05.2017 &	122 & S10 \\
7895.627 &	0.159 &	22.05.2017 &	123 & S10 \\
7896.415 &	0.291 &	22.05.2017 &	113 & S11 \\
7896.604 &	0.323 &	23.05.2017 &	154 & S11 \\
7897.601 &	0.491 &	24.05.2017 &	136 & S11 \\
7898.598 &	0.659 &	25.05.2017 &	148 & S11 \\
7899.417 &	0.797 &	25.05.2017 &	 92 & S11 \\
7899.597 &	0.828 &	26.05.2017 &	140 & S11 \\
7900.613 &	0.999 &	27.05.2017 &	 78 & S11 \\
7901.590 &	0.163 &	28.05.2017 &	144 & S11 \\
7910.418 &	0.651 &	05.06.2017 &	170 & S12 \\
7910.584 &	0.679 &	06.06.2017 &	127 & S12 \\
7911.418 &	0.820 &	06.06.2017 &	139 & S12 \\
7911.549 &	0.842 &	07.06.2017 &	123 & S12 \\
7912.610 &	0.021 &	08.06.2017 &	 95 & S12 \\
7913.419 &	0.157 &	08.06.2017 &	171 & S12 \\
7914.419 &	0.325 &	09.06.2017 &	173 & S12 \\
7914.550 &	0.347 &	10.06.2017 &	130 & S12 \\
7915.553 &	0.517 &	11.06.2017 &	106 & S12 \\
\end{supertabular}
\end{center}

\twocolumn
\begin{center}
\begin{table}[]
\caption{Observing log of PEPSI@VATT spectra of \inc from March 2017}\label{TabVAT}
\begin{tabular}{c c c c c}
  \hline
  \hline\noalign{\smallskip}
\setcounter{footnote}{0}HJD\footnote{a}&Phase\footnote{b}&  Date  & S/N$_{\rm III}$\footnote{c}& S/N$_{\rm V}\footnote{d}$\\
\hline\noalign{\smallskip}
7819.804    &	0.381	&     07.03.2017 &  50  &  80  \\
7819.959    &	0.407	&     07.03.2017 &  42  &  73  \\
7820.847    &	0.557	&     08.03.2017 &  31  &  82  \\
7820.990    &	0.581	&     08.03.2017 &  41  &  94  \\
7821.804    &	0.718	&     09.03.2017 &  54  &  96  \\
7821.963    &	0.745	&     09.03.2017 &  45  &  103  \\
7822.779    &	0.882	&     10.03.2017 &  45  &  97  \\
7822.958    &	0.912	&     10.03.2017 &  50  &  94  \\
7823.838    &	0.061	&     11.03.2017 &  38  &  97  \\
7824.008    & 0.089	&    11.03.2017 &  22  &  60  \\
7824.792    &	0.221	&     12.03.2017 &  39  &  97  \\
7824.986    &	0.254	&    12.03.2017 &  36  &  89  \\
7825.798    & 0.391	&     13.03.2017 &  45  &  88  \\
7826.003    &	0.425	&    13.03.2017 &  48  &  83  \\
7826.799    &	0.560	&     14.03.2017 &  41  &  104  \\
7826.994    &	0.593	&     14.03.2017 &  53  &  104  \\
7827.792    &	0.727	&    15.03.2017 &  48  &  90  \\
7827.945    &	0.753	&     15.03.2017 &  47  &  85  \\
\hline\noalign{\smallskip}
 \end{tabular}
\setcounter{footnote}{0}\footnote{a}{2\,450\,000+}\\
\footnote{b}{Phases computed using Eq.~1.}\\
\footnote{c}{Signal-to-noise ratio using CD\,III (blue) cross-disperser}\\
\footnote{d}{Signal-to-noise ratio using CD\,V (red) cross-disperser}
\end{table}
\end{center}


\begin{figure*}[tb]
\centering
\vspace{0.35cm}
\hspace{0.2cm}\large{S01}\hspace{5.4cm}\large{S02}\hspace{5.4cm}\large{S03}
\vspace{0.35cm}

\includegraphics[width=0.56\columnwidth]{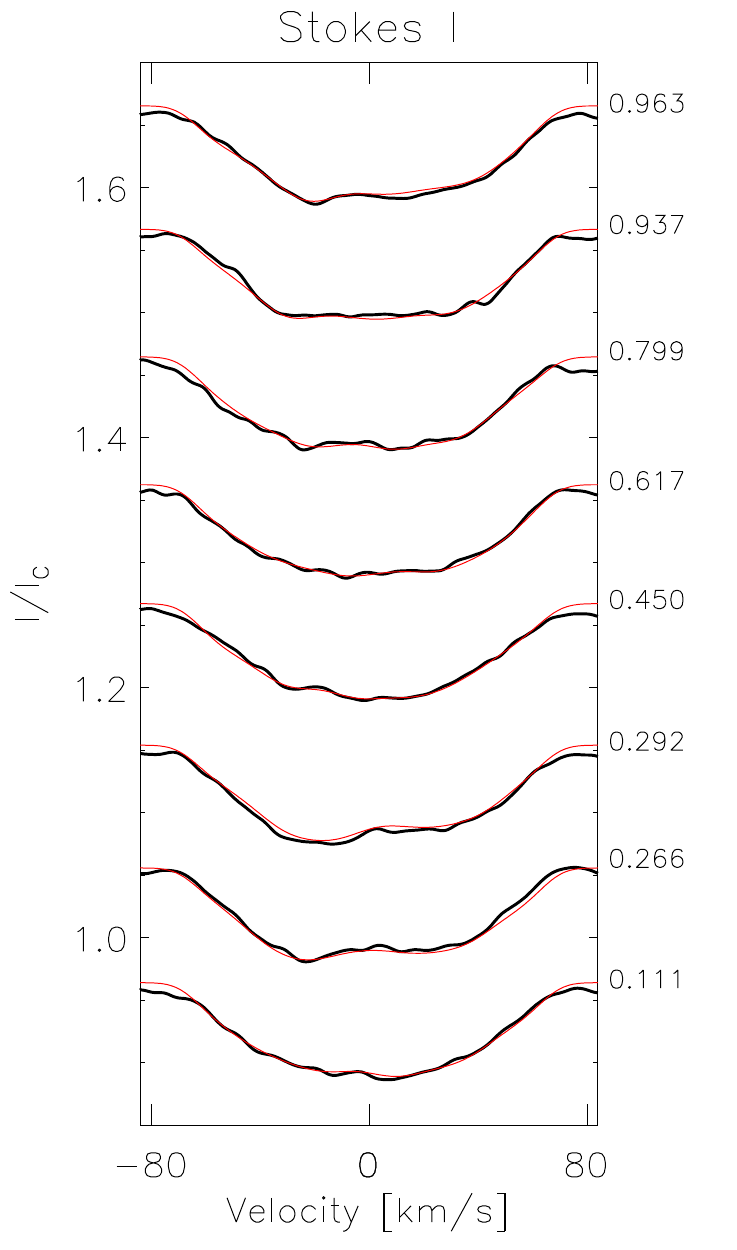}\hspace{9mm}
\includegraphics[width=0.56\columnwidth]{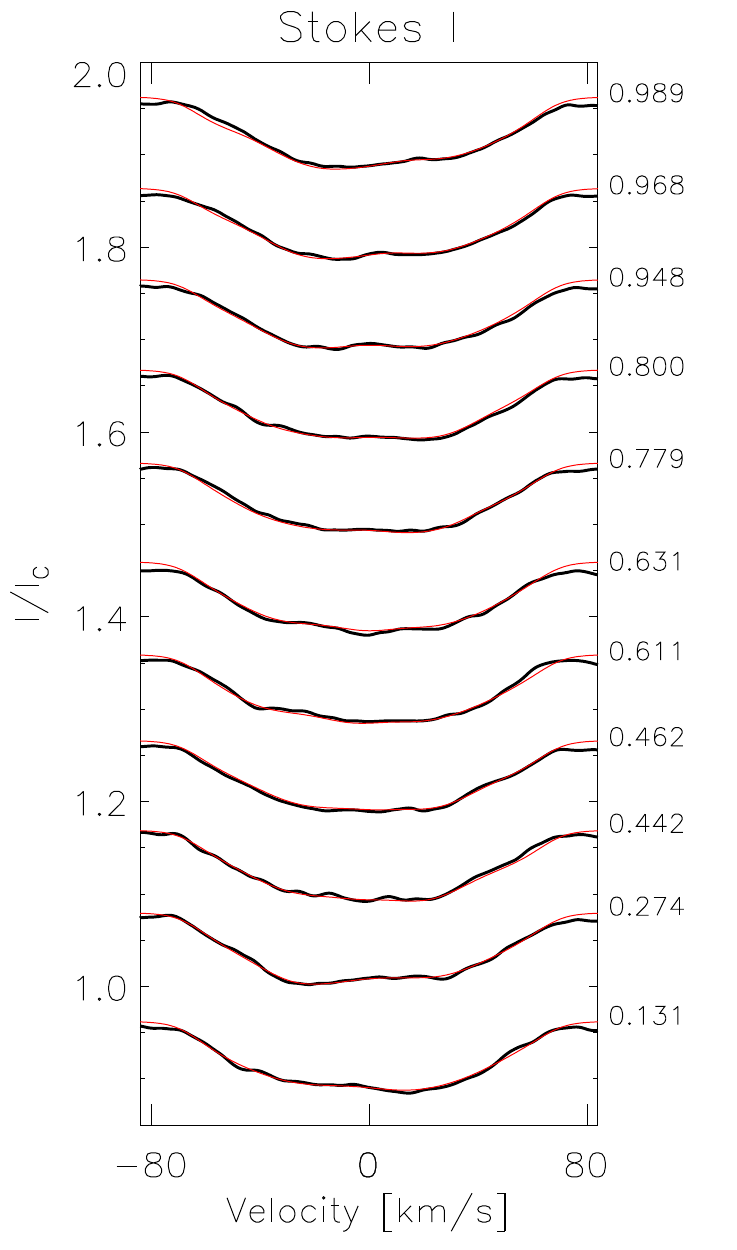}\hspace{9mm}
\includegraphics[width=0.56\columnwidth]{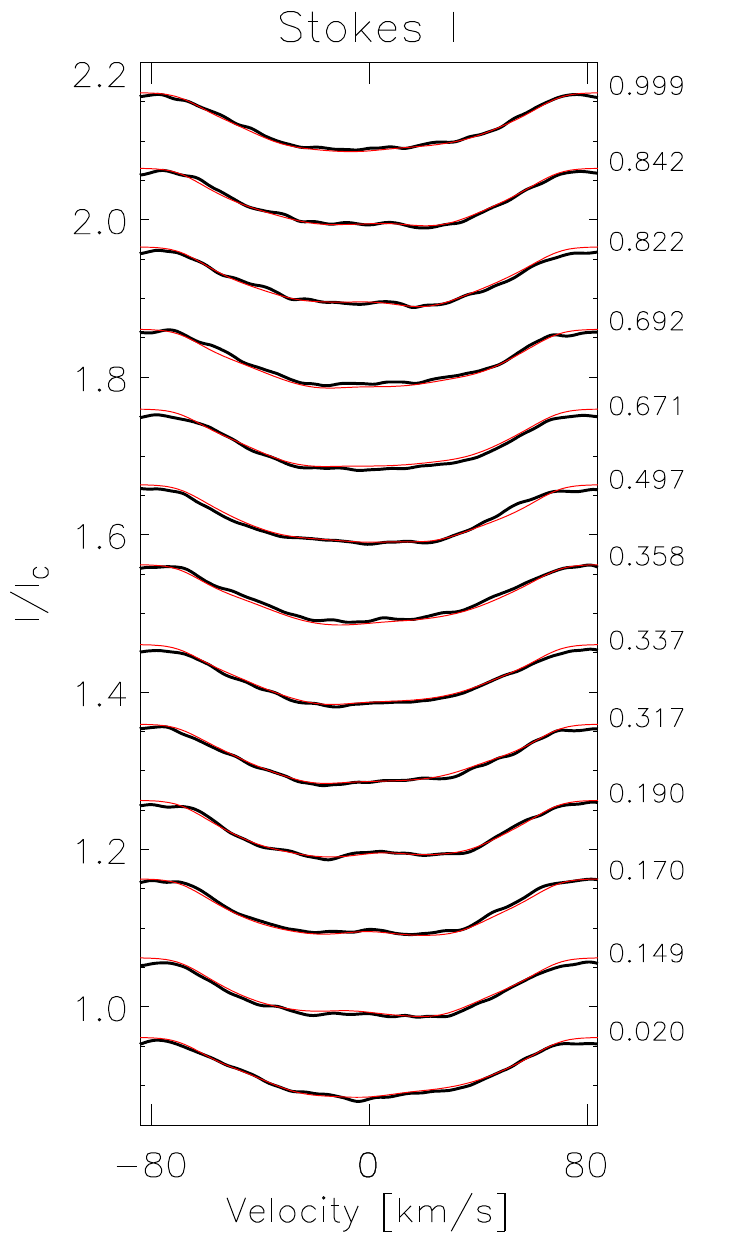}

\centering
\vspace{1.0cm}
\hspace{0.2cm}\large{S04}\hspace{5.4cm}\large{S05}\hspace{5.4cm}\large{S06}
\vspace{0.35cm}

\includegraphics[width=0.56\columnwidth]{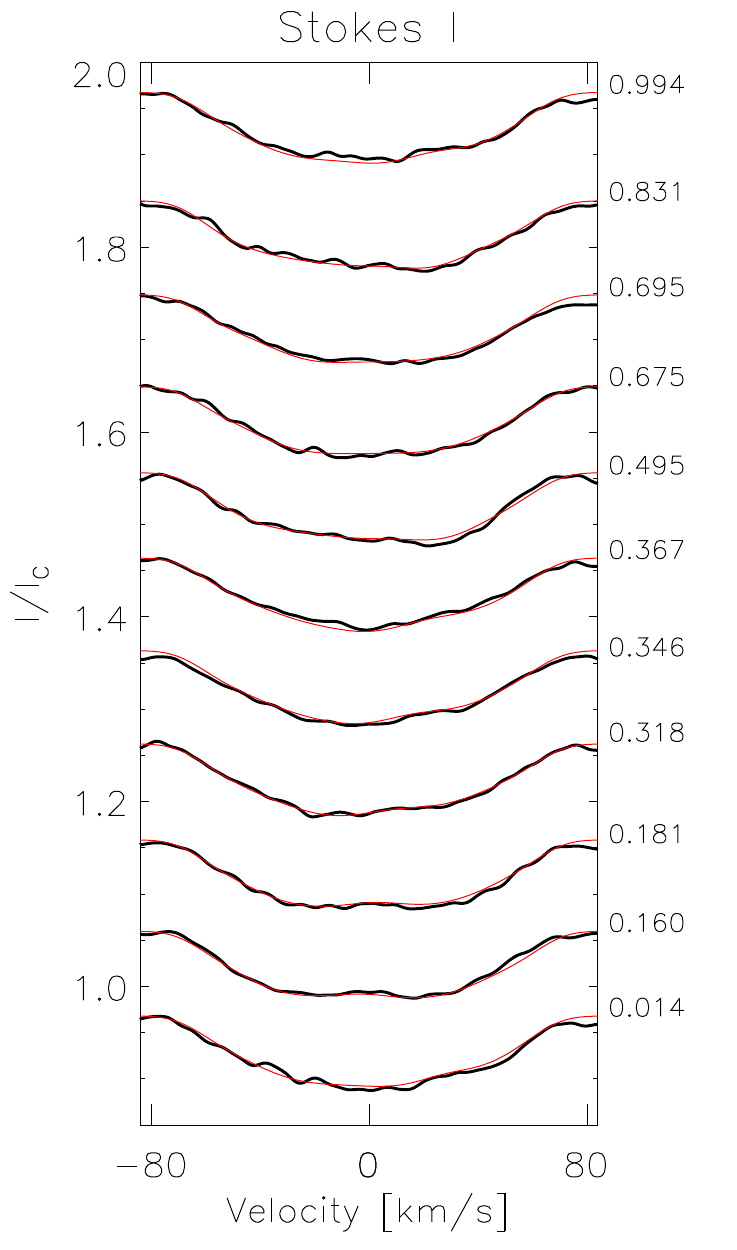}\hspace{9mm}
\includegraphics[width=0.56\columnwidth]{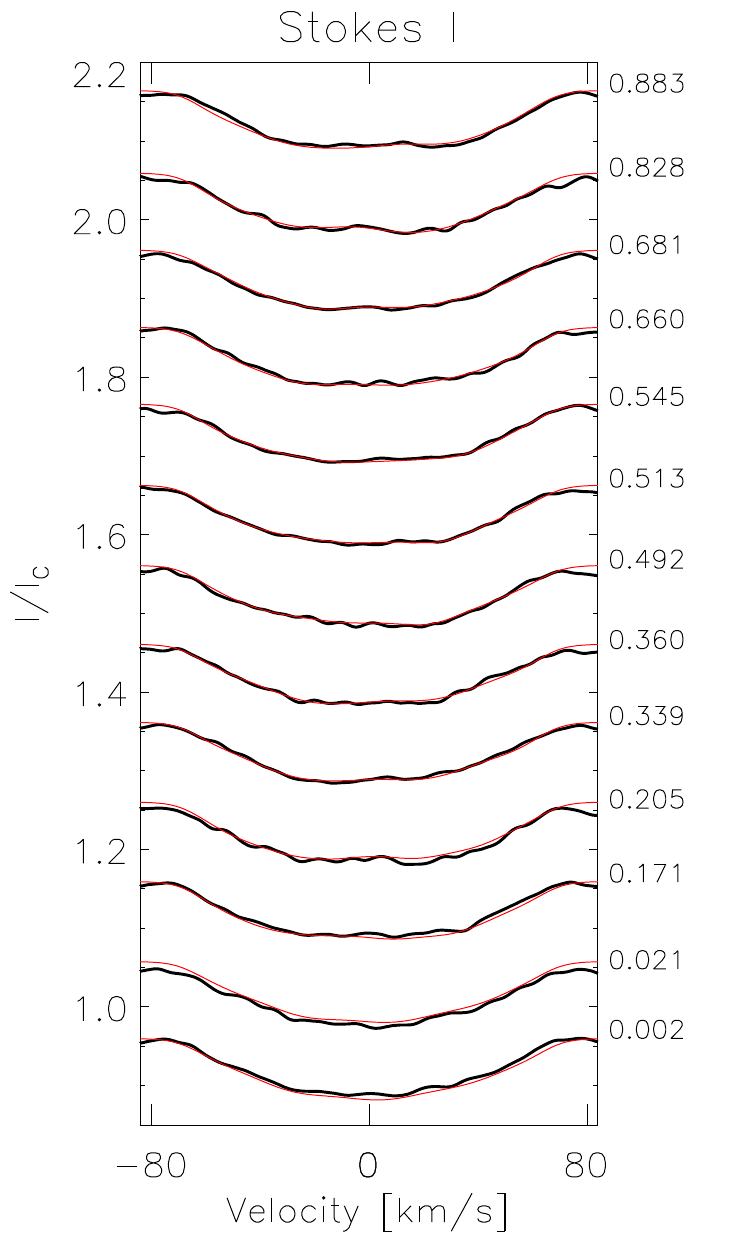}\hspace{9mm}
\includegraphics[width=0.56\columnwidth]{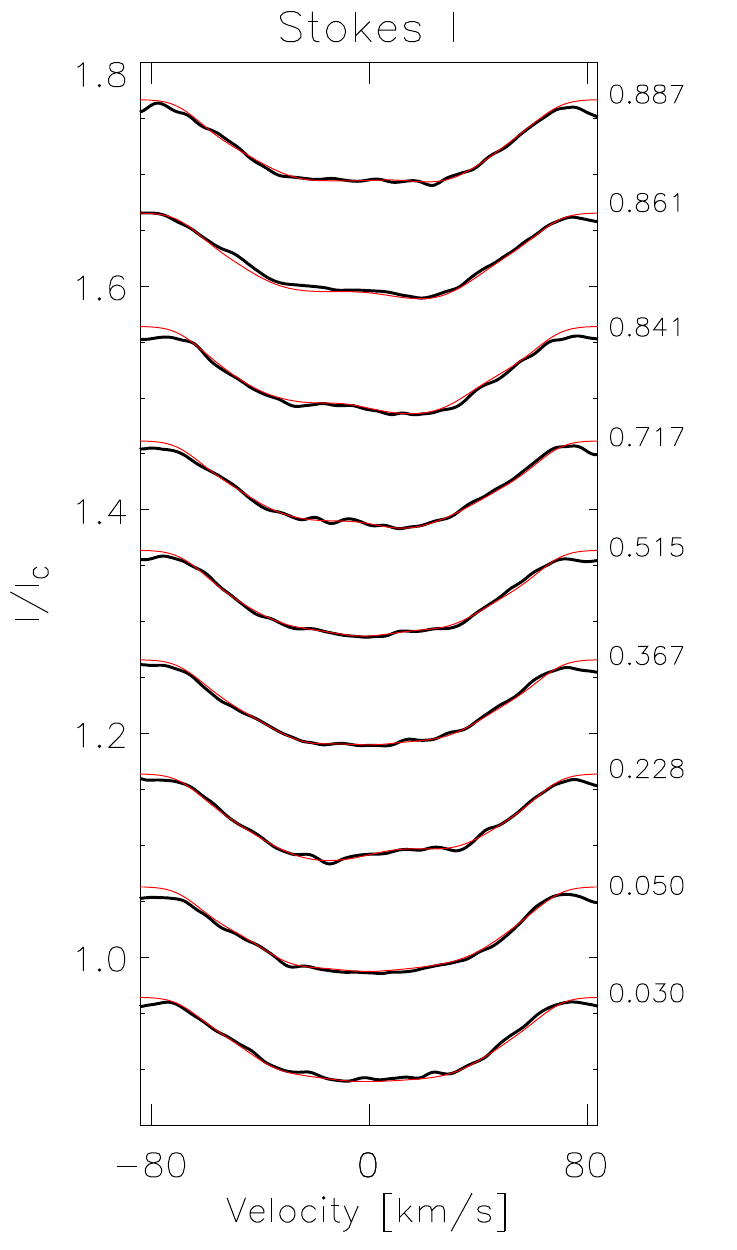}
\vspace{0.55cm}

\caption{Observed line profiles (thick black lines) and their model fits (thin red lines) for the Doppler reconstructions S01-S06 shown in Fig.~\ref{dis1}.
The phases of the individual observations are listed on the right side of the panels.}
\label{proffits1}
\end{figure*}


\begin{figure*}[tb]
\centering
\vspace{0.35cm}
\hspace{0.2cm}\large{S07}\hspace{5.4cm}\large{S08}\hspace{5.4cm}\large{S09}
\vspace{0.35cm}

\includegraphics[width=0.56\columnwidth]{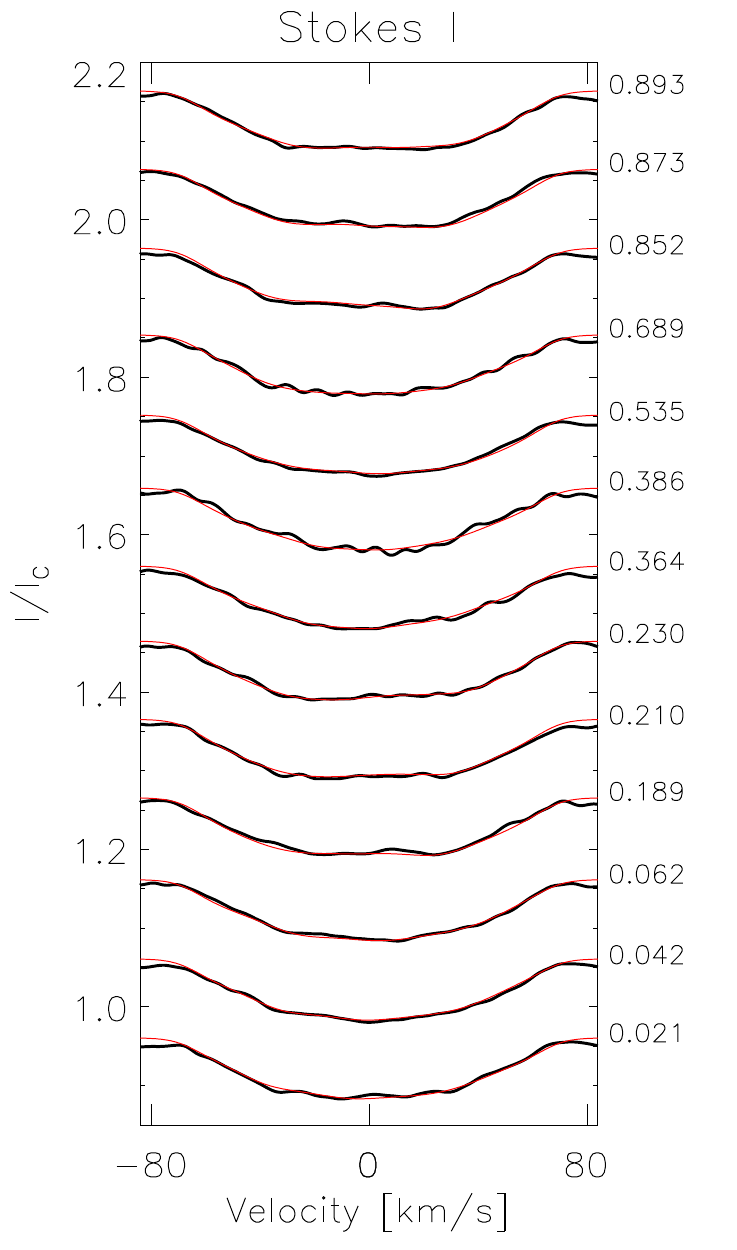}\hspace{9mm}
\includegraphics[width=0.56\columnwidth]{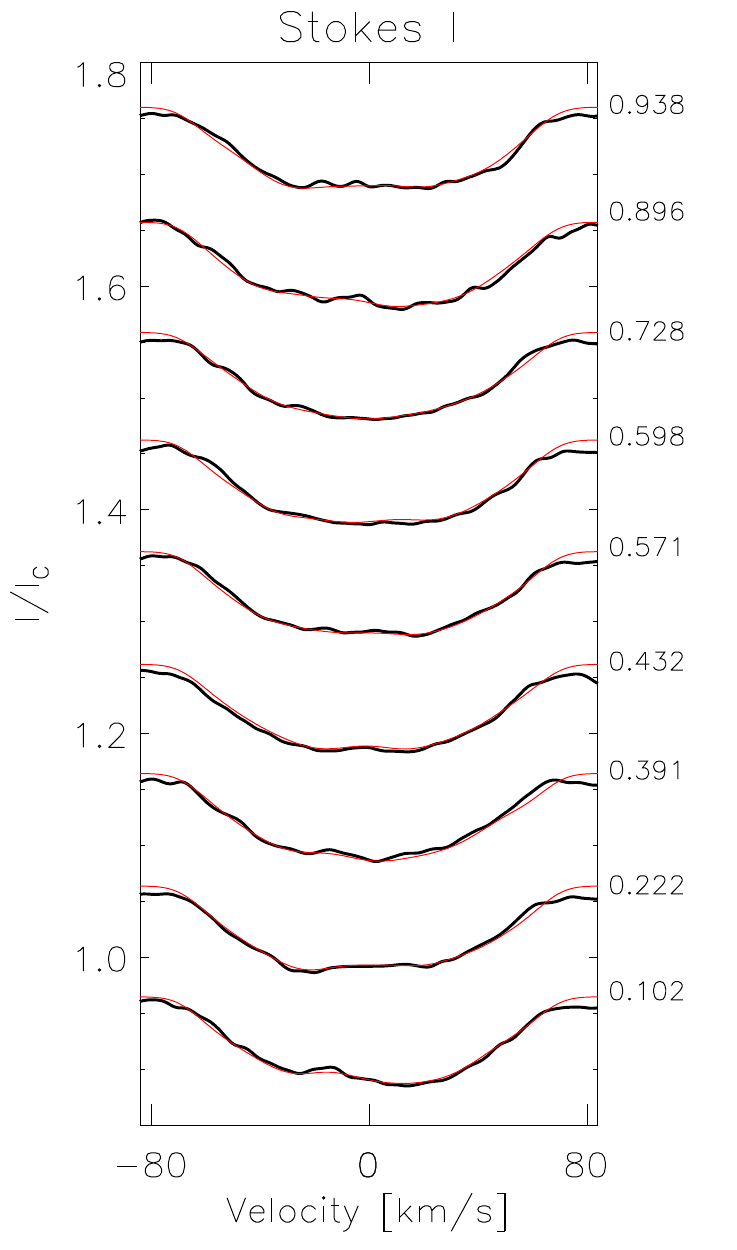}\hspace{9mm}
\includegraphics[width=0.56\columnwidth]{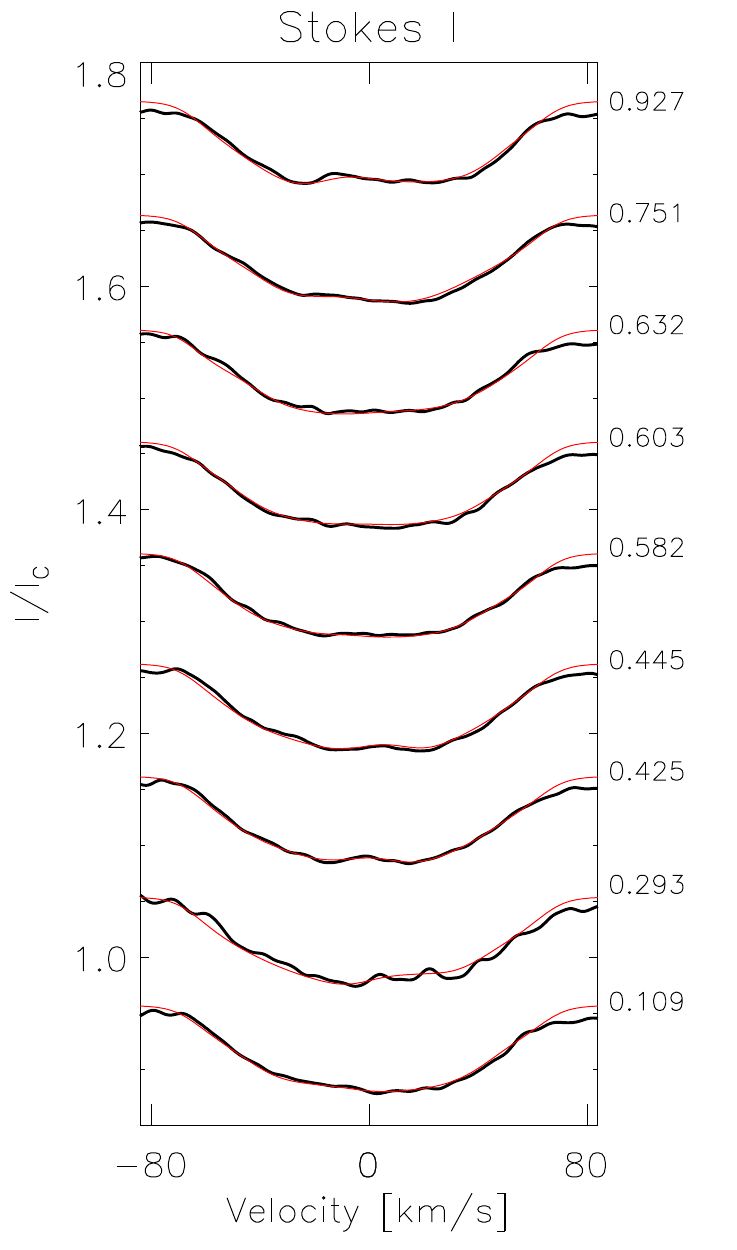}

\centering
\vspace{1.0cm}
\hspace{0.2cm}\large{S10}\hspace{5.4cm}\large{S11}\hspace{5.4cm}\large{S12}
\vspace{0.35cm}

\includegraphics[width=0.56\columnwidth]{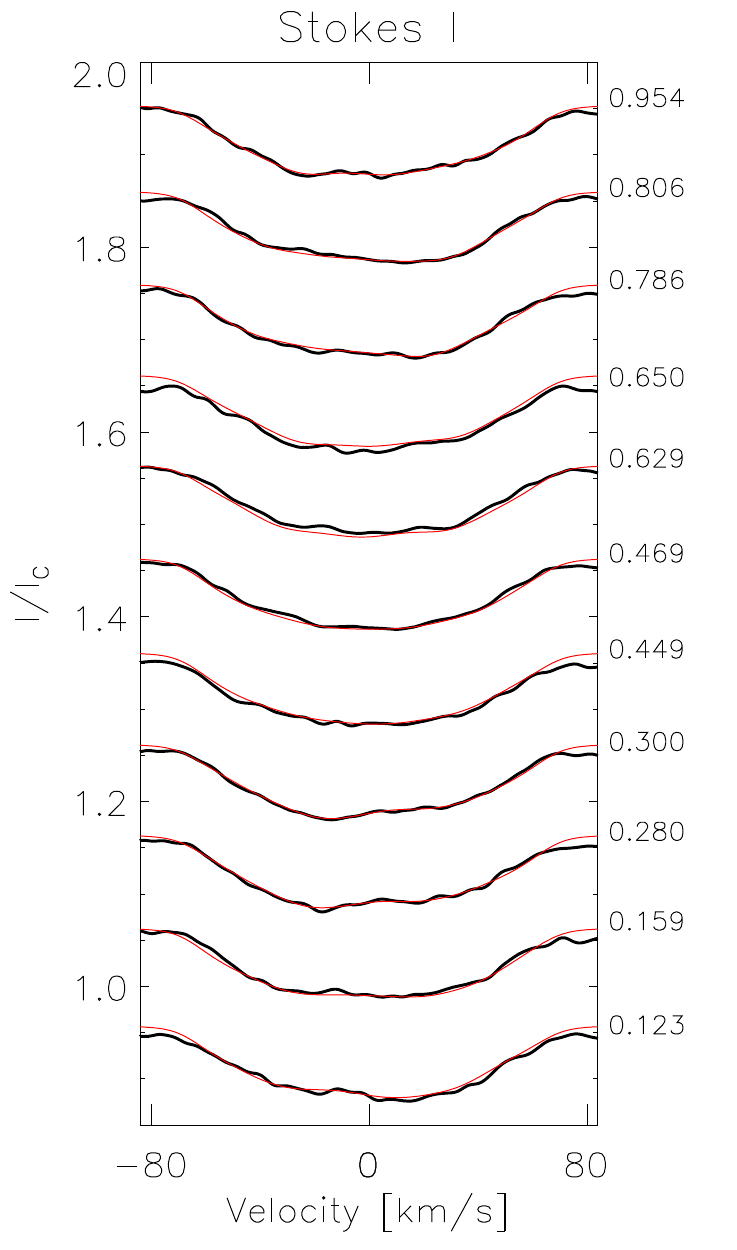}\hspace{9mm}
\includegraphics[width=0.56\columnwidth]{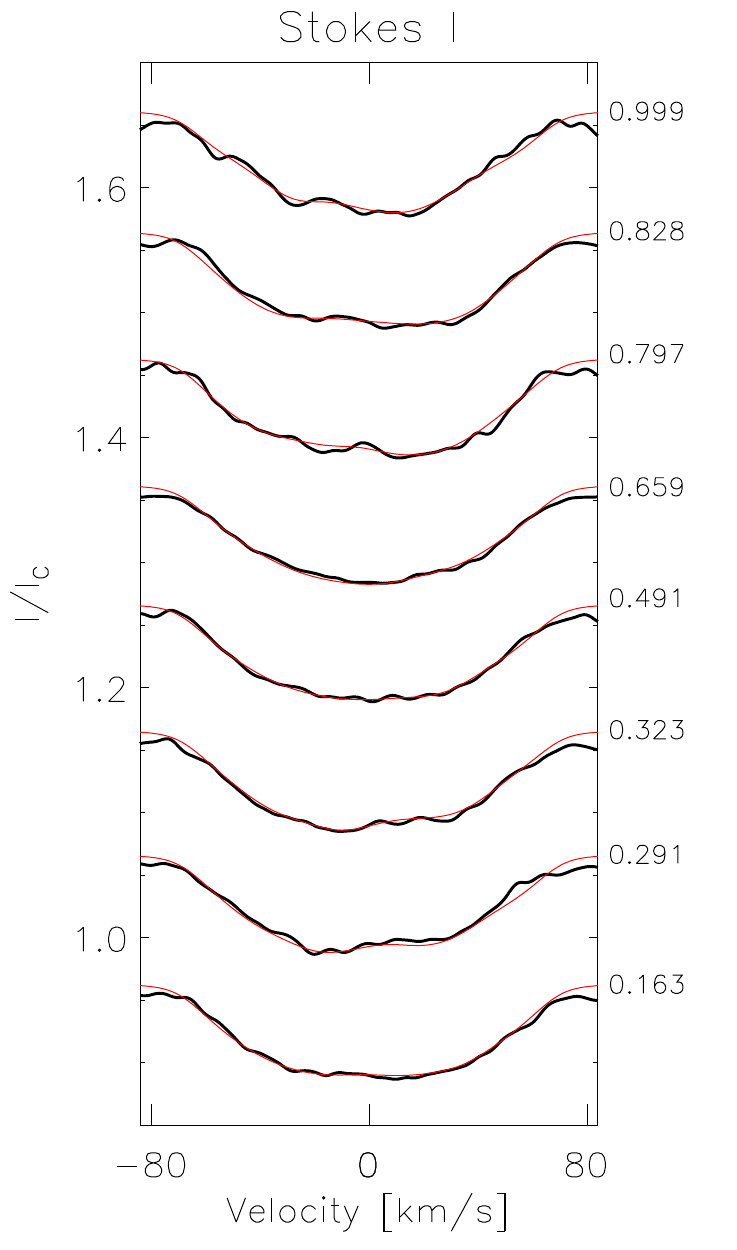}\hspace{9mm}
\includegraphics[width=0.56\columnwidth]{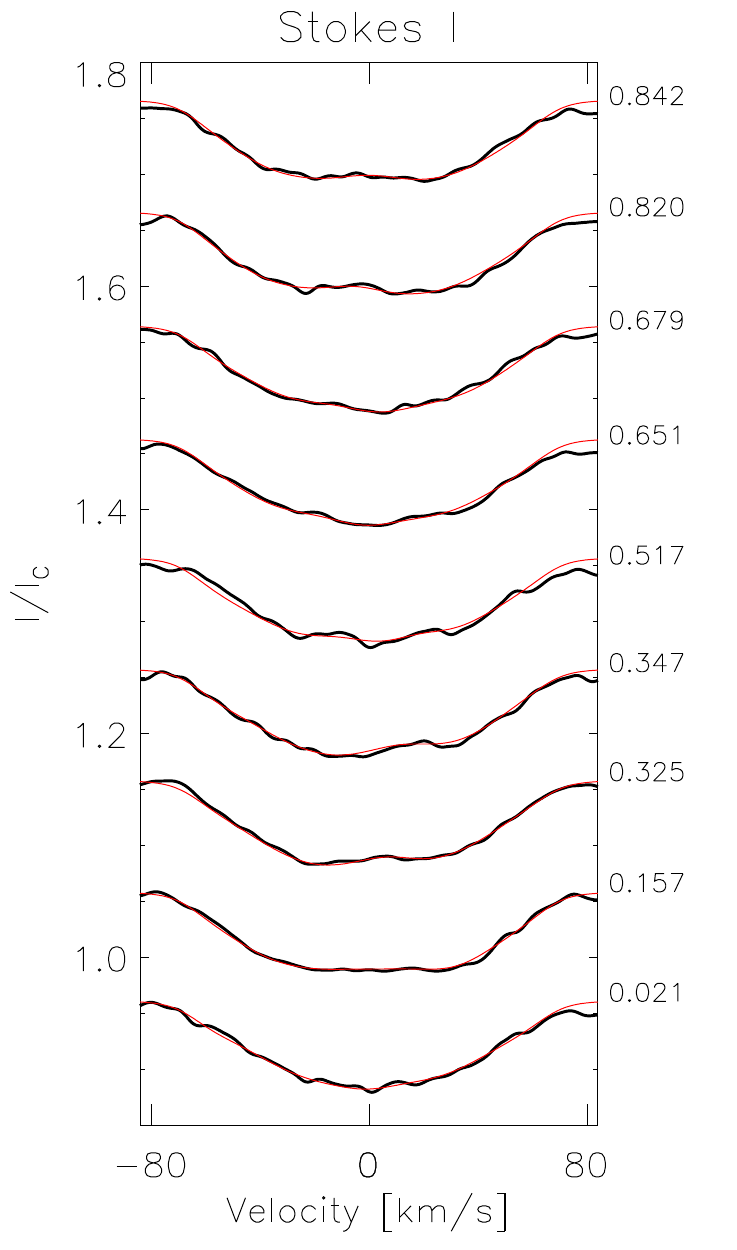}
\vspace{0.55cm}

\caption{Observed line profiles (thick black lines) and their model fits (thin red lines) for the Doppler reconstructions S07-S12 shown in Fig.~\ref{dis2}.
The phases of the individual observations are listed on the right side of the panels.}
\label{proffits2}
\end{figure*}

\begin{figure*}[tb]
\centering
\vspace{0.35cm}
\vspace{0.35cm}

\includegraphics[width=0.56\columnwidth]{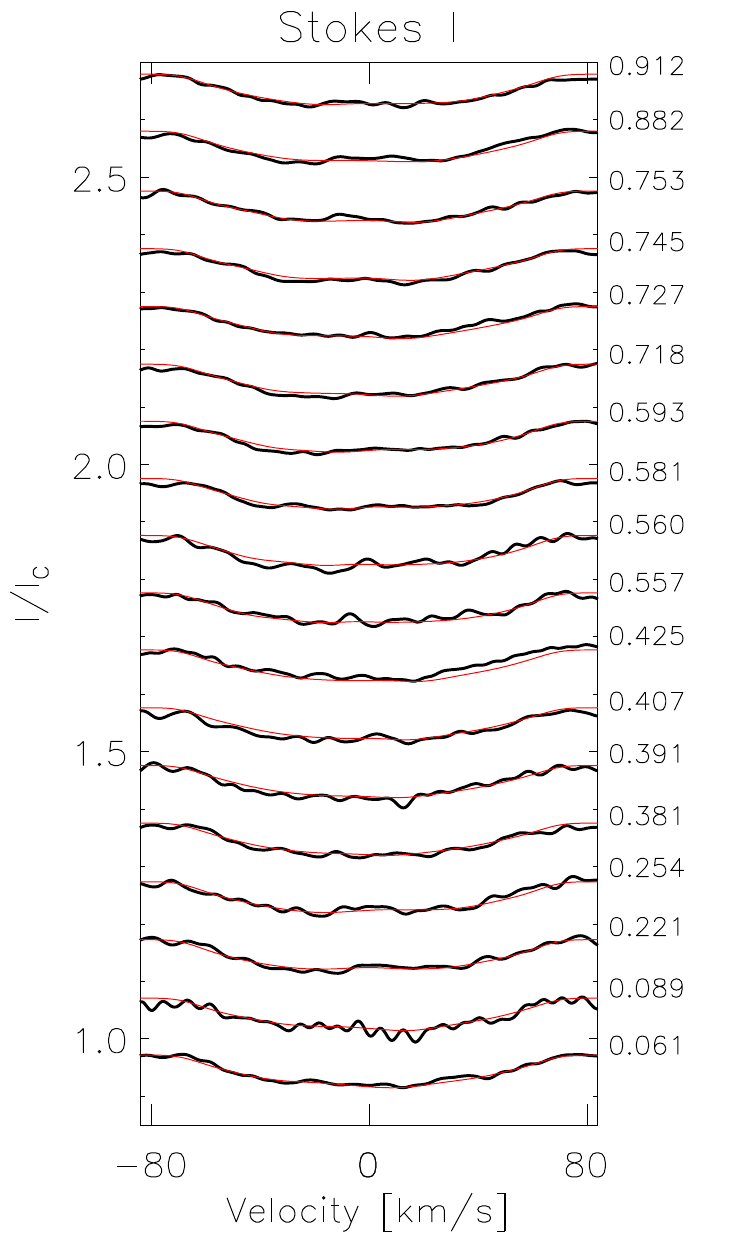}\hspace{9mm}

\caption{Observed line profiles (thick black lines) and their model fits (thin red lines) for the Doppler reconstruction applied for the PEPSI@VATT spectra shown in Fig.~\ref{dipepsi}.
The phases of the individual observations are listed on the right side of the panel.}
\label{proffits3}
\end{figure*}

\end{appendix}

\appendix

\end{document}